**High-Fidelity Electron Spin Gates for Scaling Diamond Quantum Registers**

T. Joas[1], F. Ferlemann[2,5], R. Sailer[1], P.J. Vetter[1], J. Zhang[1], R. S. Said[1], T. Teraji[3], S. Onoda[4], T. Calarco[2,5,6], G. Genov[1], M. M. Müller[2], F. Jelezko[1,7]

*[1] Institute for Quantum Optics, Ulm University, 89081 Ulm, Germany*

*[2] Peter Grünberg Institute - Quantum Control (PGI-8), Forschungszentrum Jülich GmbH, 52428 Jülich, Germany*

*[3] Research Center for Electronic and Optical Materials, National Institute for Materials Science, Tsukuba, Ibaraki 305-0044, Japan*

*[4] Takasaki Advanced Radiation Research Institute, National Institutes for Quantum Science and Technology (QST), Takasaki, Gunma 370-1292, Japan*

*[5] Institute for Theoretical Physics, University of Cologne, 50937 Köln, Germany*

*[6] Dipartimento di Fisica e Astronomia, Università di Bologna, 40127 Bologna, Italy*

*[7] Center for Integrated Quantum Science and Technology (IQST), Ulm University, 89081 Ulm, Germany*

Diamond is a promising platform for quantum information processing as it can host highly coherent qubits that could allow for the construction of large quantum registers. A prerequisite for such devices is a coherent interaction between nitrogen vacancy (NV) electron spins enabling scalable entanglement. Entanglement between dipolar-coupled NV spin pairs has been demonstrated, but with a limited fidelity and its error sources have not been characterized. Here, we design and implement a robust two-qubit gate between NV electron spins in diamond and quantify the influence of multiple error sources on the gate performance. Experimentally, we demonstrate a record gate fidelity of $F_{2q} = (96.0 \pm 2.5)$ % under ambient conditions. Our identification of the dominant errors paves the way towards NV-NV gates beyond the error correction threshold.

## I. INTRODUCTION

Quantum processors have evolved from a scientifically intriguing concept into powerful devices on the verge of solving certain computational problems efficiently[1–3]. Their success has been enabled by the increased number of controllable qubits on recent devices, which can reach three-digit numbers[3,4]. Nevertheless, many compelling computational problems require operating quantum error correction protocols on thousands of individually controllable qubits, connected through high-fidelity multi-qubit gates. Each quantum platform faces technical challenges in scaling towards this goal[5–8]. Actual devices supporting error correction have only modest size[9–12] or have not been tasked with computational problems yet[4].

Nuclear spin qubits in diamond, which are controlled by a single nitrogen-vacancy (NV) electron spin, offer gate fidelities among the leading platforms[13,14], reaching 99.94[15] % and 99.93[15] % for single-qubit and two-qubit gates, respectively, and support register sizes of up to ten nuclear spins[16,17]. Their capability to operate under ambient conditions can lower requirements on experimental infrastructure and classical control electronics, potentially allowing preservation of high-fidelity gates as the number of qubits increases. However, scaling up the register size in diamond with a single NV is restricted by the limited range of the nuclear dipolar interaction and spectral crowding, making diamond a challenging candidate for building quantum processors, capable of error correction.

To overcome these scaling limitations, photon-mediated entangling gates[18] have been implemented as a proof of concept to optically link separate nuclear registers[19,20], albeit with low entanglement rates. Complementarily, it is appealing to directly connect multiple nuclear spin registers within the same diamond by coupling their central NV spins via the longer-range electron dipolar interaction. A central capability for such a connection, electron spin entanglement, has already been achieved but with a limited fidelity[21–23]. Improved performance is possible for state-to-state transfers in NV-NV registers with optimal control microwave pulses[24]. Yet, optimization of a full two-qubit gate for arbitrary input states remains a significant challenge at room temperature as the available error metrics imply increased measurement time and higher computational overhead[15,25,26]. Furthermore, experimental implementation is often impeded by the individual, non-linear microwave response of each experimental setup.

In this work, we apply an alternative approach to optimal control for fidelity improvement, where we characterize the physical error sources. Starting with a relatively simple gate, we analyze how errors affect its dynamics via a combination of experiments and simulations. Modeling the physics of the full system of four spins (two electron + two nuclear spins) enables us to quantify gate error sources that determine the entangling gate fidelity and to mitigate their effect in a targeted manner. This allows us to apply simple, sine-envelope[27,28] microwave pulses while optimizing the entangling gate parameters (see Methods), which directly contribute to gate infidelity.

To demonstrate the effectiveness of our approach, we use randomized benchmarking to measure two-qubit gate fidelities in an electron spin quantum register in ambient conditions. We achieve a two-qubit gate fidelity of $F_{2q} = (96.0 \pm 2.5)$ %, outperforming the reported Bell state fidelities in the literature[24] – even without the previously needed optimal control pulses.

As a result of the optimization, our gate operates close to the $T_2$ decoherence limit, and we characterize the remaining coherent errors. For diamond quantum computing hardware beyond the noisy intermediate-scale quantum era (NISQ[29]), the goal will be to surpass the error correction threshold. Our model projects that this objective is attainable at room temperature, and it identifies the physical mechanisms that promise the largest enhancements in fidelity. The prolonged electron coherence time (up to $T_2 \sim 1$s[30]) at cryogenic temperature could further improve NV-NV gate fidelity.

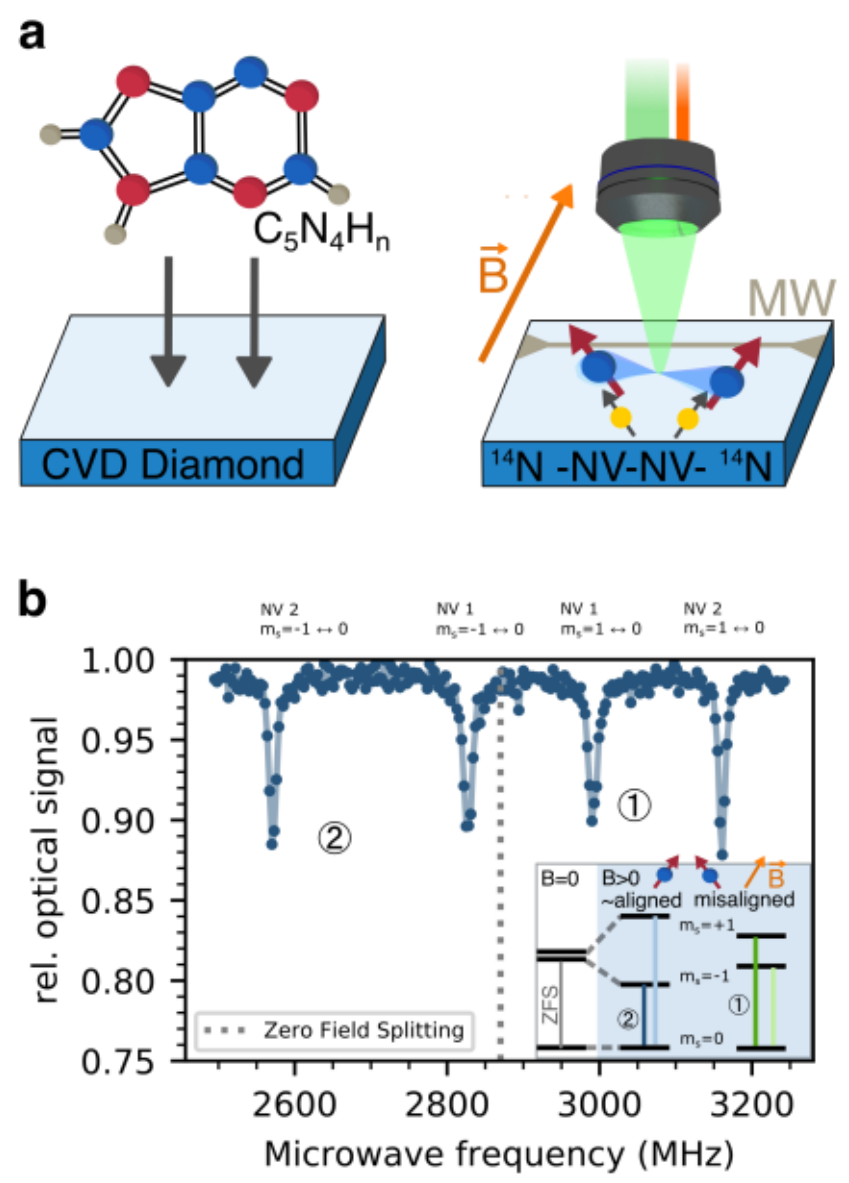

Figure 1. Diamond quantum register. (a) Left: The register is fabricated by implanting adenine molecules into an epitaxially-grown diamond layer. Right: The coupled NV system (consisting of two NV centers plus their inherent $^{14}N$ nuclear spins) is accessed optically with a confocal microscope. Spin states under a bias magnetic field $\vec{B}$ are manipulated through a microwave antenna. (b) Optically detected magnetic resonance spectrum of the register in panel (a). We employ the marked transitions 1 and 2 as "target qubit" and "control qubit". In principle, these labels are interchangeable. Inset: Sketched energy diagram of a single NV without and with (blue background) magnetic field applied. Here, the control qubit is better aligned with the magnetic field and thus shows a higher splitting in the microwave domain.

## II. RESULTS AND DISCUSSION

### a) Diamond quantum register

The quantum register for our gate is fabricated by implanting $C_5N_4H_n$ molecules from an adenine ion source[31] into a $^{12}C$ enriched, epitaxially grown diamond layer (see Fig. 1a and details in Methods section). Two of the four nitrogen atoms in the molecule can end up in close proximity ($\approx 9 \pm 4$ nm) in the diamond lattice and form two negatively charged nitrogen vacancy (NV) centers. The estimated probability of finding an NV pair in a confocal spot is ~15 %. $NV^-$ initialization and readout are achieved optically: We initialize the spin state with a green laser pulse. If required, the charge state is probed by a weak orange laser pulse exciting only the negative charge state of the NV center[32]. Note that the NV centers are too close to be optically resolvable based on their spatial separation. Hence, room temperature, state-of-the-art readout allows only to monitor the total fluorescence from both qubits upon green laser illumination. The readout signal is proportional to the summed qubit populations, but lacks information about correlations between them (see Methods)[21]. Each NV in the pair has a different orientation in the crystal lattice, resulting in spin sublevels of different energies when we apply a magnetic bias field, e.g., $|B|{\sim}100$ G (Suppl. Note 1), of appropriate orientation. A continuous-wave optically detected magnetic resonance (ODMR) experiment thus shows four lines; two below and two above the zero-field splitting energy (ZFS, Fig. 1b). We define our qubit subspace by the choice of microwave transitions shown in the inset of Fig. 1b. The difference in transition frequencies allows for single qubit rotations by applying pulses with frequencies that match the resonance of the respective qubit.

We can detect the presence of a magnetic dipolar interaction of $\nu_{dip}{\sim}0.1$ MHz between NVs with a Double Electron Electron Resonance[33] (DEER) experiment. As the coupling is smaller than the reciprocal dephasing time $1/T_2^*{\sim}0.5$ MHz, the entangling gate needs to be embedded into a decoherence-protecting dynamical decoupling pulse sequence. Extending upon previous work[21] to increase the coherence during the operation, we employ a two-qubit gate that applies robust XY8 microwave pulse sequences[34] on both NVs (see Fig. 2a). Tuning the spin flip time, $\tau_2$ between the center of the $\pi$ pulses on the second NV and the times when the first NV is refocused, we partially refocus the dipolar interaction, adjusting the effective interaction time and acquired phase. Equipped with single-qubit rotations and an entangling operation, we have a gate set that in principle allows for arbitrary two-qubit quantum computations[35].

### b) Entangling gate

Our entangling gate, labelled the $\sqrt{ZZ}$ gate is:

$$U_{\sqrt{ZZ}} = diag(1, i, i, 1) \qquad (1)$$

in the basis of the computational basis states $\{|00\rangle, |01\rangle, |10\rangle, |11\rangle\}$ and up to a global phase (see Suppl. Note 4). We obtain additional intuitive understanding from the spin evolution on the Bloch sphere. To this end, we bring the first NV ("target qubit") into a superposition state and apply our $\sqrt{ZZ}$ gate sequence. Choosing such an input state renders the experiment a dynamically decoupled DEER[36] and varying the spin flip time $\tau_2$ allows us to measure the dipolar coupling (see Fig. 2).

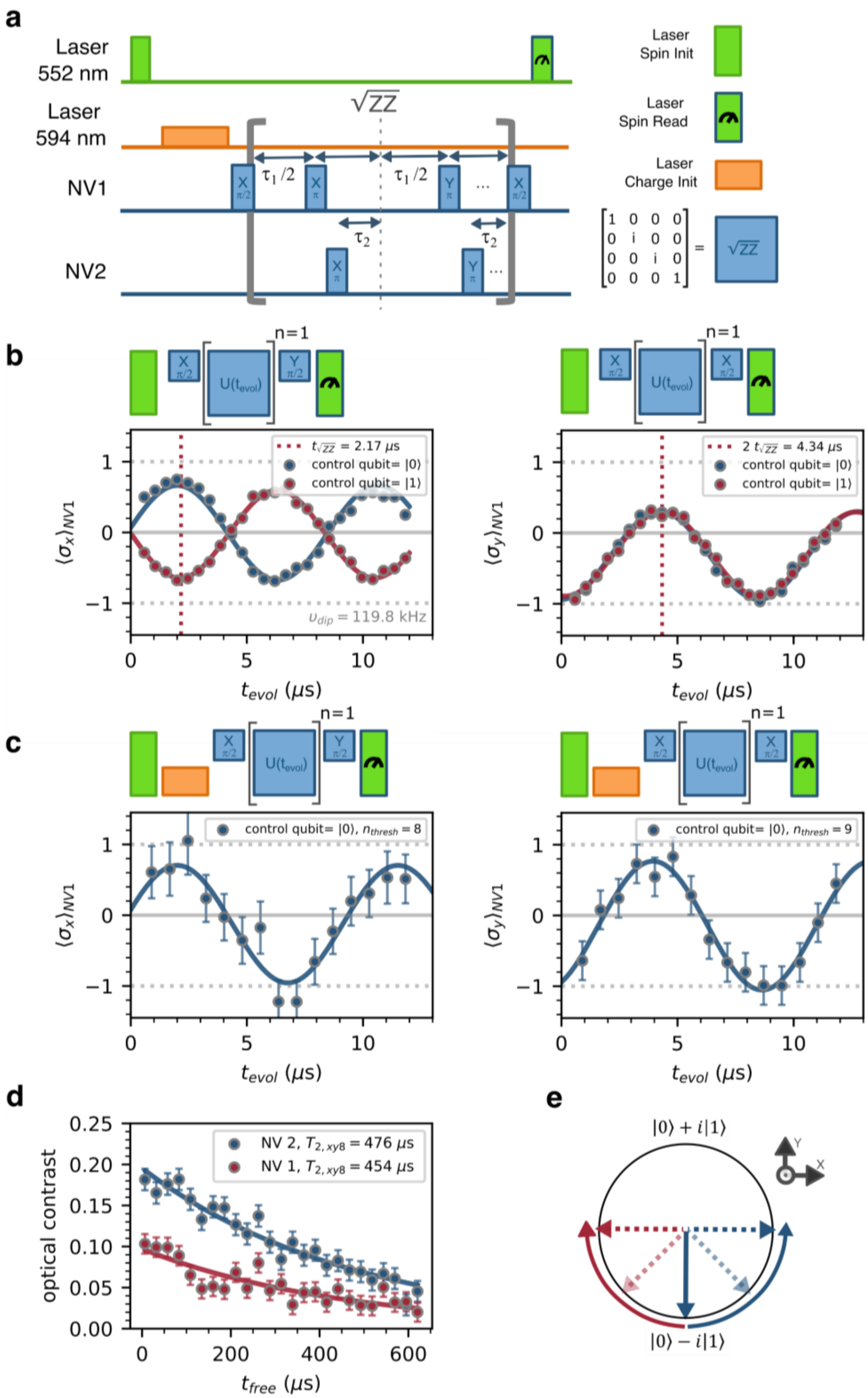

Figure 2. Entangling gate. **(a)** Microwave and laser sequence for probing the $\sqrt{ZZ}$ entangling gate. Blue: Envelope-shaped microwave pulses resonant with NV 1 (target qubit) and NV 2 (control qubit) of labeled phase (X or Y) and pulse area ($\pi/2$ or $\pi$). For each NV, $N_\pi = 8$ (phase cycled with XY8-1) $\pi$ pulses inside the grey brackets form the $\sqrt{ZZ}$ gate. For the sake of a clear figure, only two of the eight pulses per NV are drawn. The $\pi/2$ pulses around the gate are used to initialize the target qubit to the Bloch sphere equator and map the evolution under the gate onto the $\sigma_z$ readout axis. Green: 3 µs pulses of the 552 nm laser used for spin initialization and readout (blue stroke). Orange: 3.5 ms pulses of the the 594 nm laser for optional charge state initialization. The unitary matrix on the right is realized for a calibrated $\tau_1$ = 800 ns, $\tau_2 = t_{\sqrt{ZZ}}/N_\pi$, $n = 1$. **(b)&(c)** Measured evolution of the $\sigma_x$, $\sigma_y$ component of the target qubit for varying $\tau_2$ in the $\sqrt{ZZ}$ gate sequence ($t_{evol} = N_\pi n \tau_2$, here $\tau_1 = 3000$ ns, $n = 1$, $N_\pi = 8$; fitted

with $y = A_s \sin(2\pi v_{dip} + \phi_s) + y_{0,s}$, where $A_s, \phi_s, y_{0,s}$ are free parameters). The $\sqrt{ZZ}$ gate is realized after $t_{\sqrt{ZZ}}$. The target qubit is initialized and mapped onto $\sigma_z$ as given by the respective microwave sequence. Charge initialization with threshold parameter $n_{thresh} = 8, 9$ is applied for (c). **(d)** $T_2$ coherence measurements for both NVs by applying a XY8-n sequence with fixed pulse spacing $\tau_1$=800 ns sequence and varying order $n_{xy}$ (total sequence duration $t_{free} = 8n_{xy}\tau_1$) to only one of the NVs, respectively. Fixing $\tau_1$ while increasing $n_{xy}$ naturally compares to repeated application of our gate sequence that uses fixed $\tau_1$, $\tau_2$. **(e)** Idealized spin dynamic under the gate with initial $\pi/2$ pulse as in (b)&(c) on the Bloch sphere equator of the target qubit for the cases control qubit state = $|0\rangle/|1\rangle$ (blue/ red). A control qubit conditioned rotation from $\tau_2 = 0$ with no acquired phase evolves for $\tau_2 > 0$.

In the $\sigma_x$ component, we observe (Fig. 2b) a sinusoidal evolution arises as more phase is acquired through the dipolar interaction by a longer $\tau_2$ (sketch in Fig. 2e). The controlled nature of the gate that is required for a computationally complete gate set is revealed by the reversed direction of the $\sigma_x$ DEER oscillation when the control qubit is initialized into $|1\rangle$. We extract the dipolar coupling $v_{dip} = (119.8 \pm 1.0)$ kHz between the two NVs from the fitted sine oscillation frequency. Thus, we find a calibrated evolution time $t_{\sqrt{ZZ}} = N_\pi \tau_{2,\sqrt{ZZ}} = (2.17 \pm 0.02)$ µs and describe details and optimization of the $\pi$ pulse spacing $\tau_1 = 800$ ns in Suppl. Note 4. When calibrating, we minimize unwanted interactions with the inherent $^{14}$N nuclear spins[37] due to the misaligned magnetic field to the NV axes.

An important aspect of the gate performance is inferred from the sine oscillation amplitude and offset parameters (ideally: $A_s = 1, y_{0,s} = 0$). The apparent non-zero offset in Fig. 2b can be explained by the imperfect initialization of the NV charge state. If any of the NVs is neutrally charged ($NV^0$), no ZZ phase is collected during the $\sqrt{ZZ}$ gate; the output state is independent of $\tau_2$ and thus far off the target state. Applying a green (552 nm) laser pulse yields a steady state charge distribution of ~0.7 $NV^-$ and ~0.3 $NV^0$ for a single NV[38] and thus a probability of $p(NV^-, NV^-) = 0.49$ for both NVs to be in the desired negative charge state. Overall, the observed oscillation is an average over the different charge cases. This lowers the contrast of the $\sigma_x$, $\sigma_y$ oscillation in Fig. 2b. Second, the $\sigma_y$ component becomes asymmetric (offset parameter of fitted sine $y_{0,s} < 0$), as this component stays $\langle\sigma_y\rangle_{NV1} = -1$ for all $\tau_2$ given a wrong charge state and the degree of asymmetry ($|y_{0,s}|/A_s$ ) can be a direct measure of the expected charge state of the NV pair (Suppl. Note 5).

In Fig. 2c, we mitigate the charge initialization error by using a weak orange laser that employs thresholding (with a parameter $n_{thresh}$) over data sets containing all detected fluorescence photons[32]. The DEER oscillations in Fig. 2c, show improvements in contrast and asymmetry of the $\sigma_y$ component. Yet, there remains a significant deviation from full, symmetric contrast. This is in accordance with the limited charge state fidelity $F_{NV-,NV-} = (83 \pm 6)$ % measured in a separate measurement in Suppl. Fig. 4. We also show that the charge state fidelity can be improved by increasing the threshold parameter $n_{thresh}$ of the post-selection. However, stricter thresholding comes at the cost of a worsened signal-to-noise ratio. Nevertheless, higher photon collection efficiency, provided, e.g., by micro structured lenses[39], multi-laser pulse excitation[40] or doping-induced NV charging[41–43], could enhance charge state fidelity significantly.

Given the minimal gate time $1/(2v_{dip}) = 4.2$ µs and the coherence times $T_{2,XY8}$ = (454 ±58) µs, (476 ±30) µs for each NV measured in Fig. 2d, we calculate a coherence limit for the entangling gate fidelity of 99.1 %, and 98.6 % for our exact microwave gate with the actual $\tau_1$. However, our analysis in Section E reveals that, in the long coherence regime enabled by dynamical decoupling, gate errors related to the $^{14}N$ nuclear spins become significant. We note that the slight asymmetry of the applied sequence on NV 2 might not be optimal in terms of robustness[44] but such sequences have been demonstrated to be favorable for decoupling in the presence of cross-talk errors[45]. In addition, asymmetry allows us flexibility in terms of the choice of $\tau_1$ and the total gate duration.

### c) Repetitive benchmarking

Imperfections that are not related to the state evolution during a gate are commonly referred to as state preparation and measurement errors (SPAM)[46].

We use repetitive benchmarking, i.e. repeated application of the gate to a certain input state[47], to separate the entangling gate fidelity from SPAM errors. We demonstrate that this tool quantifies gate errors independently of the charge state. The latter error can be significantly reduced by charge state mitigation strategies, as described in the previous section.

Separation of the gate error is achieved by repeated application of the gate under test. After n repetitions, single qubit rotations are applied to map back to the ground state and the $\langle\sigma_z\rangle$ expectation value is measured. In the resulting decay curves, errors occurring during the gate operation are collected in the lifetime parameter of a fitted model which we convert to a pseudo error per gate (pEPG) metric (see Methods).

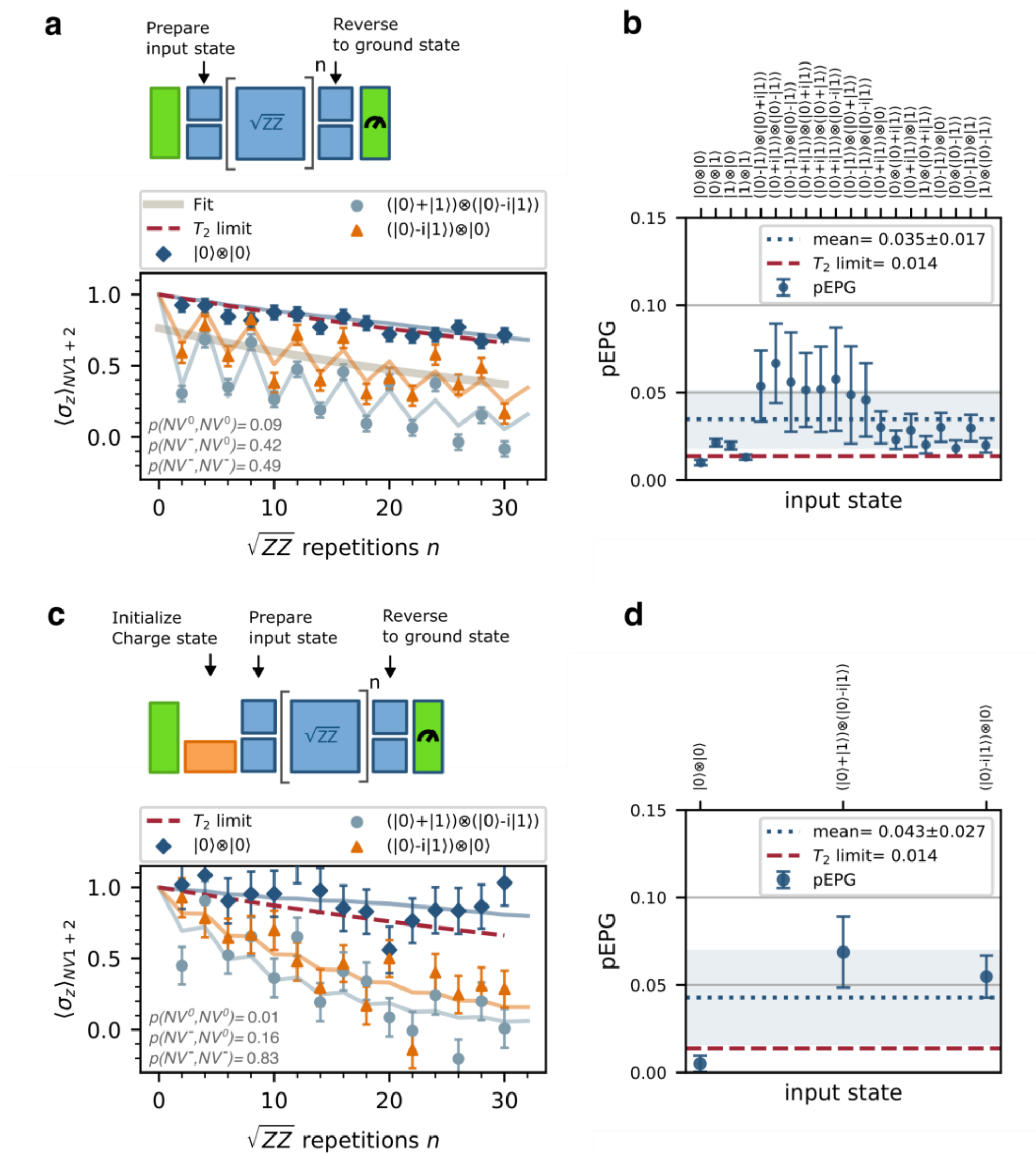

Figure 3. Repetitive benchmarking. **(a)&(c)** Measured surviving population after $n$ applications of the $\sqrt{ZZ}$ gate. For each data point, a reverse circuit is added that returns to $|00\rangle$ in the absence of gate and preparation errors. Data for exemplary input states, an exponential fit (grey line) to extract the pEPG for input state $(|0\rangle - i|1\rangle)\otimes|0\rangle$ (simplifying notation, this abbreviates $|\Psi\rangle_{NV1}\otimes|\Psi\rangle_{NV2} = \frac{|0\rangle - i|1\rangle}{\sqrt{2}}\otimes|0\rangle$ with dropped normalization throughout this work), and simulation results including the charge state behavior are shown (colored solid lines, charge state probabilities annotated). **(b)&(d)** Extracted pEPG from the data in (a)&(c) for all probed input states. The charge-state-induced modulation in the decay data described in the main text leads to the plotted $1\sigma$ fit errors that depend on the input state. Blue shading represents the standard deviation of all data points. Charge initialization is applied for (c)&(d). Mean is calculated as average over the mean pEPGs of each input state group to ensure comparability between (b)&(d).

Figure 3a shows decays for repeated $\sqrt{ZZ}$ gates for three input states. We find a mean $\overline{pEPG} = 0.035 \pm 0.017$ averaged over all tested input states (see Fig. 3b), corresponding to a pseudo gate fidelity of $pF_{2q} = 1 - \overline{pEPG} = (96.5 \pm 1.7)$ %. This is only a factor of 2.5 away from the $T_2$ limit $pEPG_{T2} = 0.014$ that we calculate from an independent coherence measurement (Fig. 2b). Thus, coherent gate errors are qualitatively small and we will later quantify the different contributions.

The input states can be grouped into three classes of pEPGs: The computational basis states show the lowest gate errors due to their magnetic noise insusceptibility. Gate errors increase when one qubit is in a superposition input state because of decoherence during the gate operation. Finally, input states with both qubits in superposition yield entangled states for gate repetitions n with mod(n,2)=1. Thus, not only both qubits will be affected by decoherence, but also some entangled states will pick up this noise more efficiently[48].

A prominent feature in repetitive benchmarking is an additional modulation with a period of *n*=4 for input states that are not the computational basis states. This behavior is a consequence of the charge state SPAM error. When one of the NVs is initialized as $NV^0$, the coupling is $\nu_{dip} = 0$. This SPAM error has no effect when the initial state is a computational basis state. However, imperfect initialization reduces the output state fidelity when any of the NVs is in a superposition state with double superposition states showing the greatest error (see Fig. 3a). We simulate the repetitive benchmarking experiment in Fig. 3a&c considering all charge state configurations (details in Method section). The experimentally observed dependence of the modulation depth on the chosen input state is well reproduced.

When adding charge state initialization to the repetitive benchmarking in Fig. 3c, the modulation is reduced and we extract a slightly elevated $\overline{pEPG} = 0.043 \pm 0.027$. Comparing this mean $\overline{pEPG}$ for the charge-initialized case in Fig. 3d with the steady state initialized data in Fig. 3b, we find agreement within the standard deviation calculated from the data of all input states.

### d) Randomized benchmarking

Randomized benchmarking[46] has emerged as a standard tool to evaluate gate performance, yielding a fidelity metric of an average computation with a priori limited insight into the nature of the gate error. In contrast to repetitive benchmarking, this fidelity can be directly compared to values obtained on different quantum hardware platforms[49–51]. We perform two-qubit randomized benchmarking in Fig. 4a&b. (see details in Methods section)

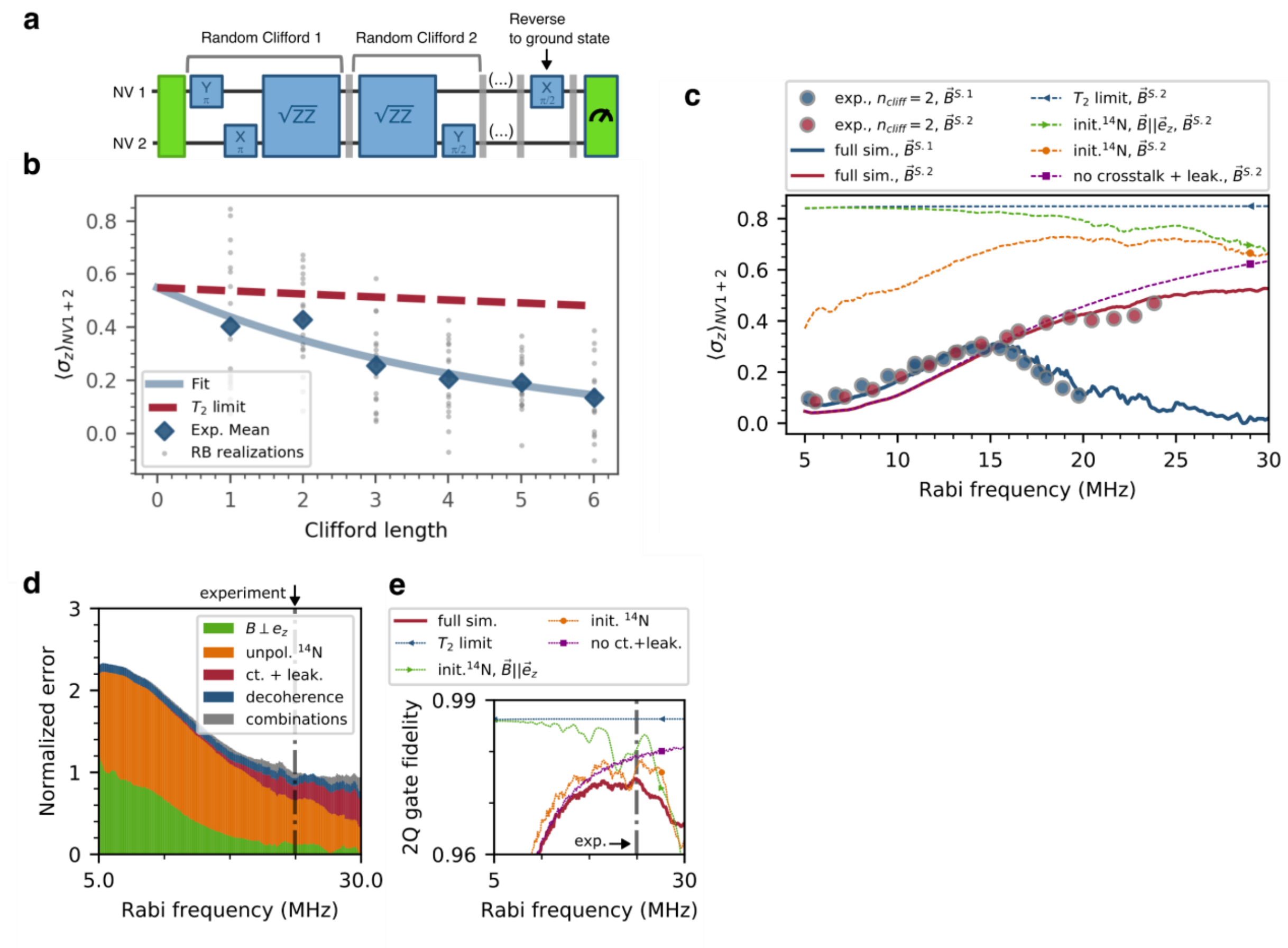

Figure 4. Randomized benchmarking and gate error analysis. **(a)** Sketched gate sequence for a single randomized benchmarking data point. Random Clifford gates are generated and reversed to the ground state. **(b)** Surviving population after randomized benchmarking in our optimized experimental setting (magnetic field setting 2, $\Omega_{rabi}/(2\pi) = 23.7$ MHz, 20 random experiments per Clifford length). From a single-exponential fit, the EPC of the gate set discussed in the main text is extracted. **(c)** Randomized benchmarking at a fixed Clifford length $n_{cliff} = 2$ for varying Rabi frequencies ($\Omega_{rabi} = \frac{\pi}{t_\pi}$ with approx. same duration $t_\pi$ of sine envelope-shaped $\pi$ pulses for both NVs). Each data point is a mean of 20 random experiments. The solid blue (red) line is a simulation at magnetic field setting 1, ~60 G (magnetic field setting 2, ~100 G), which contains a free parameter to account for the experimental SPAM errors (see Methods section). From the dashed lines, which are simulations at magnetic field 2 where the labeled error sources, i.e. crosstalk + leakage, unpolarized $^{14}$N spin, magnetic field orthogonal to the orientations of the NVs $e_z$, are turned off, we extract the relative error contributions in (d). **(d)** Error contributions to the gate set for varying Rabi frequencies, normalized to the error at the experimental $\Omega_{rabi}/(2\pi) = 23.7$ MHz, magnetic field 2. The attribution to the labeled error sources is presented in the Methods section. **(e)** Simulated gate fidelity of the $\sqrt{ZZ}$gate. The dashed lines selectively disable error sources as in (c), with only decoherence left for the blue dashed line.

In our optimized setting (magnetic field 2 in Suppl. Table 1 unless stated otherwise, $\Omega_{rabi}/(2\pi) = 23.7$ MHz) , we achieve an entangling gate fidelity of $F_{2q} = (96.0 \pm 2.5)$ % as extracted from an error per two-qubit Clifford gate $EPC = (14.9 \pm 2.7)$ % and pre-characterized average effective single-qubit errors of $F_{1q} = (99.23 \pm 0.12)$ % (see Equation 9 and Suppl. Note 3). Our entangling gate fidelity is better than the best reported NV-NV entangled state fidelity (82.4 %[24]) that was optimized by state-to-state transfer optimal control algorithms. We emphasize that the improved performance takes into account errors for arbitrary input states, in contrast to a state-to-state transfer fidelity. As we perform no charge state initialization, the measured SPAM error, represented in the amplitude parameter of the fitted exponential decay, is non-negligible. The decay's lifetime is substantially shorter than the decoherence limit ($EPC/EPC_{T2} = 5.9 \pm 1.1$). We attribute this to other, coherent errors.

### e) Gate error sources

Three error sources (explained in detail in Suppl. Note 7) mainly limit our gate set fidelity: The unpolarized $^{14}N$ nitrogen spins, the misaligned magnetic field and microwave crosstalk and leakage. We measure their influence by performing two-qubit randomized benchmarking while varying the Rabi frequency of all pulses of the gate set. To keep the measurement time viable, we determine only the surviving population $\langle\sigma_z\rangle_{NV1+2}$ for a fixed Clifford length $n_{cliff} = 2$. This allows us to extract relative error contributions, as described in the Methods section, but yields no absolute gate error metric[25]. Increasing the Rabi frequency drives the hyperfine lines due to the $^{14}N$ spin more homogenously, but is only beneficial until microwave crosstalk and leakage become limiting. For our magnetic field setting 1 (blue line in Fig. 4c, exact parameters in Suppl. Table 1) with modest frequency separation, the optimal Rabi frequency is ~(2π) 15 MHz. Our optimized magnetic field setting 2 (red line in Fig. 4c) exhibits reduced gate errors due to a large frequency separation that mitigates microwave crosstalk and leakage. Simultaneously, these errors are minimized by employing pulse envelope shaping (here: sine envelope) for all microwave pulses[27,28].

However, at higher magnetic field, state mixing for the misaligned NV 1 becomes relevant. We estimate the SPAM error introduced solely by spin mixing at the higher magnetic field setting to ~17 % by Rabi measurements on both spins in Suppl. Fig. 5a.

We employ a model (Methods section) that accurately describes the experimental randomized benchmarking results and quantifies the influence of different gate error sources. In Fig. 4c, the experimental data at Clifford length $n_{cliff} = 2$ is well reproduced by our simulation for both magnetic field settings. Turning off one error source at a time, we can extract the relative error contributions to our gate set for our optimized setting in Fig. 4d. At the experimentally chosen $\Omega_{rabi}/(2\pi) = 23.7$ MHz, the largest error contribution of 55 % is due to the unpolarized $^{14}N$ spin. The misaligned magnetic field makes up for 11 % of the observed error with crosstalk and leakage contributing 17 %, demonstrating the efficacy of our pulse envelope shaping. We note that each individual coherent error source investigated in Fig. 4 could, in principle, be corrected by adapting existing control techniques[52–54]. Their interplay and robustness within the complex four-spin Hamiltonian require further research. The relative decoherence contribution, estimated from $T_{2,XY8}$ = 454 µs, 476 µs per qubit, is 10 %.

Finally, we project the achievable entangling gate fidelity directly from our model under a realistic scenario for future experiments in Fig. 4e (97.5 % simulated at our experimental settings, incl. decoherence). Addressing all discussed coherent errors would place our gate fidelity at the coherence limit of 98.6 %. Two-qubit gate fidelities of 99.6 % that support realistic error correction protocols[55,56], are achievable by additionally extending the coherence time of the register by higher order dynamical decoupling (see Methods).

## III. CONCLUSION AND OUTLOOK

We have experimentally demonstrated the highest entangling gate fidelity of $F_{2q}$= $(96.0 \pm 2.5)$ % between solid state electron spins at room temperature. Our analysis of the gate error sources reveals that 90 % of those errors are coherent and correctable.

The primary gate error sources are the unpolarized nitrogen spin, which could be addressed by adapting one of the existing nitrogen spin polarization techniques[57–60] to our magnetic field regime, and the misaligned magnetic field that is unavoidable for NV geometries with different orientation in the crystal lattice. A perfectly aligned magnetic field, would be possible for NV quantum registers of the same orientation that are conceivable when applying strong magnetic gradient fields[61,62] (~ 1 G/nm) or leveraging different nitrogen isotopes of each NV, and it would allow a magnetic field strength compatible with straightforward nitrogen spin polarization[59]. The smaller remaining microwave errors (crosstalk and leakage) could be mitigated by optimal control[53,63–65].

We found that we incur a non-negligible SPAM error due to spin mixing in the misaligned magnetic field. Operating at cryogenic temperatures (~4 K) would significantly decrease both of the current initialization errors. First, resonant laser excitation of the sharp optical absorption lines[66] would enable fast, high fidelity charge state initialization[67,68]. Additionally, spin mixing errors could be avoided for arbitrary magnetic field geometries, as spin initialization would be possible without cycling through the NV singlet state[67] and thus independent of the off-axis magnetic field component. Furthermore, operating at low temperatures of ~4 K improves coherence times by several orders of magnitude, up to seconds[30], effectively mitigating the detrimental effect of decoherence on gate fidelities. Last, narrow optical lines would allow to distinguish the emitted photons spectrally and combination with single-shot readout techniques[67,69] could enable to measure spin correlations between the NVs.

Integrating up to four[70] coherently coupled NV centers each with nuclear spin registers of ~25 qubits[15–17] is possible with our current experimental technique. Our error analysis suggests that the gate fidelities on such a diamond quantum processor could support error-corrected operation of logical qubits. Efficient error suppression will additionally necessitate[9,12,71,72] fast, high-fidelity optical readout of nuclear[67,69] and electron spins[73,74].

## IV. METHODS

### A. Gate Fidelity Optimization

To obtain the experimental parameters realizing our highest gate fidelities, we utilize the following optimization strategy:

1. $\hat{\vec{B}}$: We align the magnetic field orientation, such that it approximately yields equally spaced microwave transition frequencies and thus reduced crosstalk and leakage. As a result, the field is closely aligned with one of the NV centers. This method also suppresses interactions with the nitrogen nuclear spin of this NV during the two-qubit gate, as the hyperfine tensor $\underline{A}$ remains diagonal.

We continue by iterative application of the following steps, each of which optimizes $F_{2q}$ with respect to one of the remaining gate parameters:

2. $\tau_1$: The undesired interaction with the nitrogen spin is strongly pronounced on the more misaligned NV. In Suppl. Note 4, we show how to select a $\tau_1$ that minimizes this undesired interaction during the $\sqrt{ZZ}$ gate. However, $\tau_1 > 2\tau_2$ reduces the effective dipolar interaction during the gate, leading to increased decoherence.
3. $|\vec{B}|$: We simulate the two-qubit gate fidelity and find the magnetic field strength that maximizes $F_{2q}$. Note the trade-off discussed in Suppl. Note 6: Higher $|B|$ can increase gate fidelity but lowers the electron spin initialization fidelity. Our final setting $\vec{B}^{S.2}$ is found in Suppl. Table 1.

4. $\Omega_{rabi}$: The same model is used to find the Rabi drive power that maximizes $F_{2q}$ as in Fig. 4e. We choose equal power for all microwave pulses in the gate set. Further improvement is possible by optimizing the power for all pulses independently (see Suppl. Fig. 6b).
5. $\tau_2$: We experimentally calibrate the phase picked up by dipolar interaction during the $\sqrt{ZZ}$ gate, which weakly depends on $\Omega_{rabi}$, as described in Suppl. Note 4.

### B. Sample

The sample used in this study is a type IIa (100) single-crystalline diamond film that is homoepitaxially grown on a high-pressure high-temperature type Ib substrate via microwave-plasma-assisted CVD[75]. The CVD film thickness is 20 µm. To suppress the effects of $^{13}C$ on the coherence properties of the NV centers, a $^{12}C$-enriched (99.95 %) high-purity (nitrogen concentration <1 ppb) diamond is used. The ion implantation process of this sample was already described elsewhere[31]. Ionized $C_5N_4H_n$ ions are extracted from adenine powder and accelerated to 65 keV kinetic energy. The implantation fluence of $10^8$ $cm^{-2}$ is achieved after 10 s. After ion implantation, the sample is annealed at 1000 °C for 2 h in a forming gas (4 % $H_2$ in Ar) to create NV centers and recover the diamond lattice. The sample is annealed in an oxygen environment at 465 °C for 4 h, followed by cleaning in a 1:1:1 mixture of $HNO_3$, $H_2SO_4$, $HClO_4$ at 200 °C under a pressure of 6-8 bar for 30 min.

For the NV pair used as a quantum register, we measured a dephasing (Ramsey) and decoherence (Hahn echo, XY8) times of $T_2^* = (2.60 \pm 0.11)$ µs; $(2.37 \pm 0.13)$ µs, $T_{2,HE} = (55 \pm 11)$ µs; $(150 \pm 9)$ µs , $T_{2,XY8} = (454 \pm 58)$ µs; $(476 \pm 30)$ µs on NV 1 and NV 2, respectively. Our pulse spacing definition of type $\frac{\pi}{2} - \frac{\tau}{2} - \pi - \frac{\tau}{2} - \frac{\pi}{2}$ with pulse spacing $\tau$ for the Hahn echo differs by a factor of two from ref. [76] to make it comparable with higher order decoupling sequences. Based on our prior work[31], we estimate the creation yield of NV pairs with similar coherence to $\approx 1.7$ %. The ratio of NV pairs to single NVs in the sample is $\approx 1{:}6.5$.

### C. Setup

We control and read the NV's spin and charge state with a standard home-built confocal microscope. Spin initialization & readout laser pulses of 3000 ns at a wavelength of 552 nm (green) are generated by a cw. laser and an acousto-optic modulator (AOM). The orange (594 nm) charge initialization pulses from a different, digitally modulated cw. laser are 3.5 ms long and of circular polarization to ensure equal ionization rates for both NVs during the charge initialization. The polarization of the green laser is linear and adjusted for nearly equal readout contrast for data in Fig. 3. For Figs. 2 and 4, the chosen polarization realized contrasts as listed in Suppl. Table 1.

All laser pulse shapes and microwave waveforms are sampled on an arbitrary waveform generator (Keysight M8195A) amplified and applied to the NV through a copper wire of 20 µm diameter placed on top of the diamond. Photoluminescence photons of the NV in the > 680 nm band are collected through an oil-immersion objective lens (Olympus, 60x, NA 1.35), counted by an avalanche photodiode (APD Excelitas SPCM) during a 330 ns gating window at the beginning of the spin readout pulse[77] and digitized by a counting card (FAST ComTec MCS6). The experiment is controlled by a custom measurement software (Qudi[78]) that features a software interface to quantum algorithms generated by Qiskit[79].

To convert experimentally measured fluorescence to polarization $\langle\sigma_z\rangle_{NV1+2}$, we perform every experiment twice with additional $\pi$ pulses on both NVs. The difference $\Delta_{alt}$ of such alternating experiments is then normalized to the optical contrast $I_{|00\rangle} - I_{|11\rangle}$ between the fluorescence in state $|00\rangle$ and $|11\rangle$. Our measurement relates to the readout of a register state $\rho$ as $\langle\sigma_z\rangle_{NV1+2} = \Delta_{alt}/(I_{|00\rangle} - I_{|11\rangle}) = R(\rho)$ with the readout outcome R as defined in Suppl. Note 2.[21]

### D. Simulations

For the simulation results of Fig. 4, we use the following Hamiltonian to describe the system of two NV centers in their orbital ground-state $^3A_2$ :

$$H(t) = H^0_{NV1} + H^0_{NV2} + H^{\mathrm{int}}_{12} + H^{\mathrm{mw}}(t) \qquad (2)$$

where $H^0_{NV1}$ and $H^0_{NV2}$ are the single NV-Hamiltonians, $H^{\mathrm{int}}_{12}$ the dipole-dipole coupling Hamiltonian and $H^{\mathrm{mw}}$ the microwave Hamiltonian, that is in general time dependent. The single NV center Hamiltonians[80]

$$H^0 = DS_z^2 + \vec{\omega}_e \cdot \vec{S} + QI_z^2 + \vec{\omega}_n \cdot \vec{I} + \vec{S}\underline{A}\vec{I} \qquad (3)$$

$$\text{with } \vec{\omega}_{e/n} \equiv -\gamma_{e/n}\vec{B}$$

contain the electron zero-field splitting $D$, the nuclear quadrupole moment $Q$, the Zeeman splitting due to the static magnetic field $\vec{B}$, the gyromagnetic ratio of the electron/nuclear spin in angular frequency units, $\gamma_{e/n}$, and the hyperfine coupling to the $^{14}$N nuclear spin via the tensor $\underline{A}$. Both the electron spin (initialized in its ground state) and the $^{14}$N nucleus (initialized in a maximally mixed state) are triplets: $S = I = 1$. The two NV centers have distinct crystal axes (c.f. Suppl. Fig 1a), causing misalignment between the magnetic field and the NV axes and thus a logical basis that is given by rotated eigenstates.

For simulating the dynamics of the system, we use an effective dipole-dipole interaction (along the quantization axes) in the $H_{12}^{\text{int}}$ term that is described in detail in Suppl. Note 2 and list the employed estimates of the system parameters, such as the magnetic field and interaction strength in Suppl. Note 1. Decoherence is accounted for by applying a factor $(1 - EPC_{T2})$ in Fig. 4c&d, $(1 - EPG_{T2})$ in Fig. 4e and $(1 - EPG_{T2,1q})$ for the single-qubit gates in Suppl. Fig. 6b. These factors, calculated in Methods section F, yield approximately constant decoherence contributions to the fidelities in the simulated Rabi frequency range.

In order to investigate the effect of the charge state initialization SPAM error, we perform a numerical simulation, which takes into account the probabilities that each of the NV centers is initially prepared in either the $NV^-$, or $NV^0$ charge state. This simulation employs the simplified Hamiltonian described in Suppl. Note 4 and does not take into account the effect of decoherence. Thus, we include a phenomenological single exponential decay of the observed signal in the charge state simulations in Fig. 3a&c, which is calibrated using the measured pEPGs of Fig. 3b&d.

### E. Repetitive Benchmarking

To convert the measured repetitive benchmarking decays into (pseudo) gate errors, we assume a single exponential decay model that is known to well describe usual $T_{2,XY8}$ measurements on single NVs[81]. We express the decay in terms of the gate repetition number $n$ as:

$$y = y_0 + a * \exp(-n/N_d) = y_0 + a * p^n \qquad (4)$$

where we used the identity $p = \exp\left(-\frac{1}{N_d}\right) = (1 - pEPG)$ to convert the gate decay parameters $N_d, p$ into a gate error $pEPG$. The remaining free fit parameters are the amplitude a and the offset $y_0$.

Similarly, we obtain the $T_2$ limit decay curve in Fig. 3 by calculating the polarization loss per applied two-qubit gate of length $t_{gate} = N_\pi \tau_1$ from the mean of the measured $T_{2,XY8}$ of both NVs in Fig. 2b:

$$pEPG_{T2} = 1 - \exp\left(-t_{gate}/((T_{2,XY8,NV1} + T_{2,XY8,NV2})/2)\right) \qquad (5)$$

We note that the pEPG derived here from the lifetime parameter of an exponential fit is a useful tool for easily implementable benchmarking, but the results are difficult to compare among experiments (which motivates our labeling as “pseudo” EPG). We empirically find a single exponential decay, which is also the decay model widely used in $T_2$ measurements[82]. While repetitive benchmarking is similar to $T_2$ experiments - except for the additional evolution by the dipolar coupling - the exact nature of the decay depends on the environment of the probed spins[81,83].

### F. Randomized Benchmarking

We use the same single exponential model of Equation 4 to obtain a decay parameter p for the observed randomized benchmarking decays. If the gate-dependency of the errors is small enough, the measured EPC becomes a good estimate of the average gate set infidelity[84].

$$EPC = \frac{2^{\text{n}} - 1}{2^{\text{n}}}(1 - p) \approx \frac{\sum_i (1 - F_{avg}(\tilde{C}_i, C_i))}{\sum_i 1} \qquad (6)$$

where n is the number of qubits, $\{\tilde{C}_i\}$ are the Clifford gates and $\{C_i\}$ the respective ideal ones (to account for the possibility of gate-dependent errors[84]), and

$$F_{avg}(\tilde{C}_i, C_i) = \int d\psi \, Tr(\tilde{C}_i[|\psi\rangle\langle\psi|] C_i[|\psi\rangle\langle\psi|]) \qquad (7)$$

is the overlap between the actual process of the $i$-th Clifford $\tilde{C}_i$ and the ideal $C_i$. Here, $F_{avg}$ is averaged over all possible initial pure states $\psi$.

Our experiments are generated from the Qiskit software package using the gate set $\{\pi_x, (\pi/2)_x, \pi_y, (\pi/2)_y, C_1NOT_2\}$ and 20 random experiments per Clifford. While the single-qubit rotations are easily realized on our NV hardware by microwave pulses with appropriate phases, we need to express $C_1NOT_2$ in terms of our available $\sqrt{ZZ}$ gate and multiplication with single-qubit unitaries:

$$C_1NOT_2 = \pi_{x,1} * (\pi/2)_{-y,2} * \sqrt{ZZ} * (\pi/2)_{x,1} * (\pi/2)_{y,2} * (\pi/2)_{y,1} * (\pi/2)_{-x,2} * (\pi/2)_{x,1} \quad (8)$$

After running Qiskit's automatic transpiler optimization ("Optimize1qGatesDecomposition") to reduce the number of single-qubit rotations in the gate string, we end up with a randomized benchmarking experiment that contains on average $GPC_{1q} = 10.5$ single-qubit $(\pi_x, (\pi/2)_x, \pi_y, (\pi/2)_y)$ and $GPC_{2q} = 1.8$ two-qubit $(\sqrt{ZZ})$ native gates per Clifford gate. The measured EPC, but not $F_{2q}$, could be further improved by finding a more efficient gate set with smaller $GPC_{1q}$.

Two-qubit randomized benchmarking yields an EPC that is an average error over the whole gate set. For estimating the error of our $\sqrt{ZZ}$ gate, we use the EPC definition[79]:

$$EPC = 1 - \prod_i (1 - EPG_i)^{GPC_i} = 1 - \left(1 - EPG_{2q}\right)^{GPC_{2q}} \left(1 - EPG_{1q}\right)^{GPC_{1q}} \quad (9)$$

where we collected all single-qubit gate errors in an average gate error $EPG_{1q}$. Solving for $EPG_{2q}$ allows us to estimate the two-qubit gate error from the measured two-qubit randomized benchmarking (yielding EPC) and the pre-characterized average effective single-qubit error per Clifford $EPC_{1q} = 1 - \left(1 - EPG_{1q}\right)^{GPC_{1q}}$ (data in Suppl. Fig. 2).

Our randomized benchmarking uses the single exponential model in Equation 4 as a decay model and yields results with low statistical uncertainty, if the tail of the decay curve is well captured. Because of limitations of the applicable microwave power, capturing the decay tail is not always possible in our experiment. For the two-qubit case, we thus determine the offset parameter $y_0$ of the decay separately first, by applying a high number of intentionally miscalibrated pulses, and fix the decay offset parameter. This amplifies gate errors and thus gives a value that is equally representing the high error limit. We consistently observe $y_0 < 0.001$ in our experiments.

We observe that the randomized benchmarking decay lifetime in Fig. 4b is shorter than expected from repetitive benchmarking (factor $EPC/EPC_{T2} = 5.9 \pm 1.1$, $\overline{pEPG}/pEPG_{T2} = 2.5 \pm 1.2$, coherence limit $EPC_{T2}$ derived in Methods section). We attribute this to the fact that in repetitive benchmarking, coherent errors are well refocused by the symmetric timing of the decoupling $\pi$ pulses in the $\sqrt{ZZ}$ gate sequence. In combination with random single-qubit Clifford gates in randomized benchmarking, this decoupling works less efficiently.

From simulations we find that a single exponential decay with offset $y_0 = 0$ is well describing the behavior for experimentally accessible numbers of two-qubit Clifford gates. For $n_{cliff} \gg 10$, accumulated crosstalk and leakage become more relevant, which can cause a slower, 2nd exponential decay[85] and depolarization into equal populations of the three $m_s = 0, \pm 1$ sublevels.

### 1. *Coherence limit*

For the $T_2$ coherence limit in Fig. 4, we first estimate the error caused by decoherence per two-qubit gate assuming the same single exponential decay as in the repetitive benchmarking case, thus $EPG_{T2,2q} = pEPG_{T2} = 1.4$ % (Equation 5). The decoherence errors from single-qubit gates are much smaller, as they feature shorter gate lengths. Hence, we roughly estimate $EPG_{T2,1q}$ by inserting the average single-qubit gate length into Equation 5. Knowing the average number of gates per Clifford, we obtain the coherence limit per Clifford $EPC_{T2}$ from Equation 9.

Higher order decoupling should enable prolonging $T_2$ towards the longer $2T_1$ limit. We thus estimate in Suppl. Note 8 an achievable coherence limit for the $\sqrt{ZZ}$ gate of $F_{2q} = 99.6$ % at room temperature.

### *Relative error contributions*

We extract the relative error contribution in randomized benchmarking (Fig. 4d) from the simulation as follows: First, we simulate the $\langle\sigma_z\rangle_{NV1+2}$ outcome of the experiment at $n_{cliff} = 2$ with all (coherent) error effects and multiply this value ($z_{sim}$) with the known $EPC_{T2}$ to account for the decoherence of a single average Clifford gate:

$$z_{sim,T_2} = z_{sim}(1 - EPC_{T2}) \tag{10}$$

Using the single exponential decay in Equation 4, we convert the simulated outcome of the randomized benchmarking experiment to an average fidelity parameter $p$:

$$p_{full,T_2} = \sqrt[n_{cliff}]{(z_{sim,T_2} - y_0)/a} \tag{11}$$

where we take the SPAM parameters $a, y_o$ from the measurement in Fig. 4a.

The deviation of our simulation (in terms of p) including all errors from a perfect, gate error free experiment with $p$=1 is

$$\Delta_{full,T_2} = 1 - p_{full,T_2}$$
$$\Delta_{Er_i,T_2} = 1 - p_{Er_i,T_2} \tag{12}$$

Similarly, we obtain average fidelity parameters $p_{Er_i,T_2}$ and deviations $\Delta_{Er_i,T_2}$ for turning off specific error sources $Er_i$. Then, the relative contribution $c_r(Er_i)$ is given by the difference between full simulation and the improved result without this error source $p_{Er_i,T_2}$ divided by the swing of the full experiment:

$$c_r(Er_i) = \frac{p_{Er_i,T_2} - p_{full,T_2}}{\Delta_{\mathrm{full,T_2}}} = 1 - \Delta_{Er_i,T_2}/\Delta_{full,T_2} \tag{13}$$

For example, when the error $Er_i$ is turned off and the result $p_{Er_i,T_2}$ does not change, the error contribution is $c_r(Er_i) = 0$.

Ideally, we would quantify the error introduced by an off-axis magnetic field by comparing to the application of parallel magnetic fields of different strength on both NVs in simulation. However, the exact transition frequencies that are asymmetrically distributed around the ZFS can only be recovered with a misaligned magnetic field in the simulation. Instead, we note that the experimental outcome on the electron spin with polarized nitrogen spin is equivalent to the case where the hyperfine interaction is turned off completely, because no population transferring terms $A_{ij}S_i \otimes I_j, i \neq j$ exist in $\underline{A}$ for a perfectly aligned magnetic field. Hence, we can attribute the additional error that occurs for the polarized nitrogen spin case when $A_{ij}S_i \otimes I_j, i \neq j$ terms are present, solely to the misaligned magnetic field:

$$c_r(\vec{B} \perp \vec{e}_z) = \mathrm{c_r}(HFS) - \mathrm{c}_r(\text{unpol.}_^{14}\text{N}) \tag{14}$$

Finally, we can turn off all known error sources in the simulation simultaneously - crosstalk, leakage and the hyperfine interaction - and calculate the difference to an experiment with only decoherence.

$$c_r(residual) = (1 - c_r(T_2)) - \mathrm{c_r}(HFS, ct. + leak.) \tag{15}$$

These small residual errors scale with the microwave power in a range of $c_r(residual) = [0.9\% - 0.09\%]$ for the studied Rabi frequencies between 5-30 MHz. Thus, we attribute them mainly to unwanted dipolar interaction occurring during the microwave pulses.

Some part of the gate infidelity cannot be attributed to a single error. This cumulative effect increases in parallel with crosstalk and leakage but has a small contribution (7 %) at the experimental setting.

The alternating readout described above for conversion from fluorescence to spin state can reject some of the leakage error. Although we verified in simulation that this effect is small for our magnetic field setting 2, we base Fig. 4d on a simulation without the additional $\pi$ pulses in order to not underestimate leakage.

Our simulation of randomized benchmarking in Fig. 4c does not account for experimental SPAM errors like the imperfect charge state initialization or mixing of the initialized spin state. Such SPAM errors do not change the decay parameter p in randomized benchmarking, but will be represented in the amplitude fit parameters $y_0, a$. Hence, we use a free parameter that is multiplied onto our simulation result $\langle\sigma_z\rangle_{NV1+2}$ and which is determined by least square minimization of the difference to the experimental data.

### G. Error bars

Error bars on experimental fluorescence data are estimated from the photon shot noise as $1/\sqrt{n_{phot}}$ and are error propagated to the spin state domain. Errors on parameters extracted from fits are $1\sigma$ uncertainties.

## V. DATA AVAILABILITY STATEMENT

The datasets in this study are available from the corresponding author on reasonable request. Processed data for the plots is publicly available[86].

## VI. ACKNOWLEDGMENTS

We thank Michael Olney-Fraser, Thomas Unden, Philipp Neumann and Alex Retzker for helpful discussions. T.J. acknowledges support by the Foundation of German Business (sdw). This work was funded by the German Federal Ministry of Research (BMBF) through future cluster QSENS to Project No. 03ZU2110DB and projects DE-Brill (No. 13N16207), SPINNING (No. 13N16210, No. 13N16209, and No. 13N16215), CoGeQ (No. 13N16101), DIAQNOS (No. 13N16463), quNV2.0 (No. 13N16707), Quamapolis (No. 13N15375), Quantenrepeater.Net (QR.N), EXTRASENS (No. 13N16935), and QuantumHiFi (No. 16KIS1593); DLR via Project QUASIMODO (No. 50WM2170); and Deutsche Forschungsgemeinschaft (DFG) via Project No. 386028944, No. 445243414, No. 387073854, No. 491245864, No. 546850640, and No. 554644981. This project has also received funding from the European Union's HORIZON Europe program via projects QuMicro (No. 101046911), SPINUS (No. 101135699), C-QuENS (No. 101135359), QCIRCLE (No. 14 101059999), and FLORIN (No. 101086142); European Research Council (ERC) via Synergy grant HyperQ (No. 856432); and CarlZeiss-Stiftung via the Center of Integrated Quantum Science and technology (IQST) and project Utrasens-Vir (No. P2022-06-007). Additional support was provided by JSPS KAKENHI (No. 21H04646 and No. 24K01286), JST Moonshot R&D Grant No. JPMJMS2062, MEXT Q LEAP Grant No. JPMXS0118067395, JST Adopting Sustainable Partnerships for Innovative Research Ecosystem (ASPIRE) via Grant No. JPMJAP24C1.

## VII. AUTHOR CONTRIBUTION

T.J., P.J.V., M.M.M. and F.J. conceived the experiment. T.T., S.O. fabricated, T.J., R.S., S.O. prepared and characterized the sample. T.J. wrote the code for the experiment and performed the measurements for the presented experimental data. F.F. and G.G. developed and analyzed the model and performed the simulations. T.J., F.F. and G.G. analyzed the data and wrote the manuscript. R.S.S., G.G., J.Z., S.O., T.C., M.M.M. and F.J. supervised the project. All authors reviewed and contributed to the manuscript.

## VIII. COMPETING INTERESTS

The authors declare no competing interests.

## X. SUPPLEMENTARY MATERIAL

### S.I. Magnetic Field Parameters

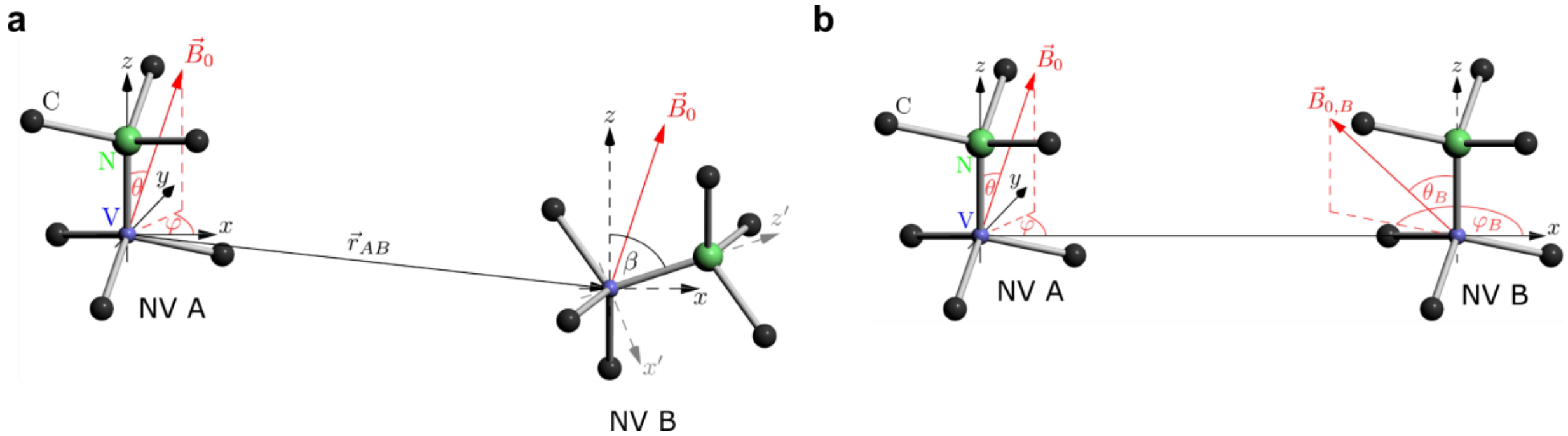

Suppl. Figure 1. Geometry of the two-NV quantum register. (a) Physical geometry of coupled NVs A and B in the crystal lattice, not to scale. The order in the gate sequence (labeled NV 1 and NV 2) is given along with geometry parameters in Suppl. Table 1. (b) Effective picture of the coupling. In this case, each NV is subject to a (unphysical) different magnetic field vector and the dipolar coupling is of zz type.

Most data presented is measured at the optimized magnetic field settings 2 of Suppl. Table 1. There, we use an aligned magnetic field on the control qubit (NV 2) for minimal interaction with the nitrogen nuclear spins. On the misaligned target qubit, the interaction during the entangling gate is much pronounced, but we can use the $\tau_1$ degree of freedom to mitigate the issue (see Suppl. Note 4). In Fig. 4b, we also use the un-optimized setting 1 to illustrate the strong influence microwave crosstalk and leakage can have.

To obtain the magnetic field geometries, we determine the microwave transition frequencies by an ODMR measurement. For each magnetic field setting, the ODMR of two NV centers yields a set of four transition frequencies, where the lowest and highest frequencies form a pair as well as the two frequencies in between, which are associated to a magnetic field of higher misalignment with the NV center axis. Given these transition frequencies and the zero-field splitting $D$, one is able to calculate the absolute magnetic field value $\omega_{e,i}/(2\pi)$ (in frequency units) and misalignment with each NV center using[87]:

$$\left(\frac{\omega_{e,i}}{2\pi}\right)^2 = \frac{1}{3}\left(f_{1,i}^2 + f_{2,i}^2 - f_{1,i}f_{2,i} - \left(\frac{D_i}{2\pi}\right)^2\right) - \left(\frac{E_i}{2\pi}\right)^2 \tag{S1}$$

$$H_i = \frac{7\left(\frac{D_i}{2\pi}\right)^3 + 2(f_{1,i}+f_{2,i})\left(2(f_{1,i}^2+f_{2,i}^2) - 5f_{1,i}f_{2,i} - 9\left(\frac{E_i}{2\pi}\right)^2\right) - 3\left(\frac{D_i}{2\pi}\right)\left(f_{1,i}^2+f_{2,i}^2 - f_{1,i}f_{2,i} + 9\left(\frac{E_i}{2\pi}\right)^2\right)}{9\left(f_{1,i}^2+f_{2,i}^2 - f_{1,i}f_{2,i} - \left(\frac{D_i}{2\pi}\right)^2 - 3\left(\frac{E_i}{2\pi}\right)^2\right)} \tag{S2}$$

$$\cos(2\theta_i) = \frac{H_i - E_i/(2\pi)\cos(2\varphi_i)}{D_i/(2\pi) - E_i/(2\pi)\cos(2\varphi_i)} \tag{S3}$$

where we defined $H_i$ for a clear notation, $f_{1,i}$ and $f_{2,i}$ are the mentioned pair of transition frequencies, $\theta_i$ the misalignment angle between the axis of the considered NV center and the magnetic field, $\varphi_i$ the azimuthal angle (in the plane perpendicular to the NV axis) and $i \in A, B$ the NV center to which the quantities are assigned. Note that the geometry in Suppl. Fig 1a is defined in terms of the physical NVs A and B, that can differ from their order in the gate sequence (NV 1 and NV 2).

Crystal strain can be divided into transversal strain $E_i$, and axial strain, which we model by including it into the zero-field parameters $D_i$[88]. Distinguishing both components from ODMR-like experimental data is not straightforward if considering a system of two NVs. Thus, we use the following methodology to determine $D_i, E_i$, making use of the assumption that the (absolute value of the) external magnetic field is constant for the closely located NVs:

As $E_i$ is typically small ($\ll$ 1 MHz) in bulk samples[88–90] we are setting it to zero. For the magnetic field setting 2, we observe that Suppl. Equations S1-S3 do not lead to a physical solution when assuming a typical zero-field splitting at room temperature of $D/(2\pi) = 2870$ MHz[77]. Thus, we obtain the solution $D_A/(2\pi) = 2865.42$ MHz that minimizes the absolute value of the axial strain on NV A and yields a valid magnetic field vector. Under the constraint that the absolute magnetic field is the same for both NV centers, this leads to an axial strain for NV B of $D_B/(2\pi) = 2867.27$ MHz. Our results for $D_A, D_B$ are not significantly altered if a (pessimistic estimate) transversal strain of $E/(2\pi) \sim 1$ MHz [91] is introduced, justifying or initial assumption of $E = 0$.

Using the same $D_A, D_B$, obtained from the analysis of magnetic field setting 2, in the lower magnetic field setting 1, we observe a small difference in the absolute magnetic field of both NVs, $|\omega_{e,A} - \omega_{e,B}|/\gamma_e = 1.1$ G.

We simplify our notation and use $\theta \equiv \theta_A$ and $\varphi \equiv \varphi_A$ in the following. Given the previously determined magnetic field misalignment $\theta_i$, we can calculate the remaining angle $\varphi$ in Suppl. Fig 1a by making use of simple geometry. The angle between the direction of the magnetic field $\hat{\vec{\omega}}_e$ and the axis of NV center B, $\hat{\vec{n}}_B = (\sin\beta, 0, \cos\beta)^T$, is the misalignment angle $\theta_B$ :

$$\hat{\vec{\omega}}_e = \hat{\vec{B}} \quad = (\cos\varphi \sin\theta, \sin\varphi \sin\theta, \cos\theta)^T$$

$$\cos\theta_B = \hat{\vec{\omega}}_e^T \cdot \hat{\vec{n}}_B = \cos\varphi \sin\theta \sin\beta + \cos\theta \cos\beta \tag{S4}$$

Note that we can find $\varphi$ up to a relative minus sign that cannot be determined in a system of two NV centers only, due to its symmetry. At the same time, as we cannot differentiate between the signs, we do not expect that the dynamics are affected by the choice of the sign.

Similarly, the geometry is not uniquely defined for $\theta$, with two solutions $\theta$ and $\theta_i' = \pi - \theta_i$ for $\theta \in [0, \pi)$.

Furthermore, the angle between the NV center axes can take two possible values $\beta' = 109.47°$ and $\beta = 180° - \beta' = 70.53°$ due to the crystal structure of diamond.

Suppl. Table 1. Parameters as chosen and estimated in the simulation. The computational basis of the qubit system (defined as $|0\rangle = |g\rangle, |1\rangle = |e_{1/2}\rangle$) employs the new spin eigenstates (due to the tilted quantization axis) that are superpositions of the $m_s = |0\rangle, \pm|1\rangle$ electron spin states, with $|g\rangle$ referring to the spin ground state and $|e_{1/2}\rangle$ to the excited spin eigenstates (lower/higher transition frequency than ZFS).

| Magnetic field setting | **Setting 1** $|\vec{B}^{S.1}| \sim 60\text{G}$ | | **Setting 2** $|\vec{B}^{S.2}| \sim 100\text{G}$ | |
|---|---|---|---|---|
| NV center (gate sequence/physical order) | NV 1 / NV A | NV 2 / NV B | NV 2 / NV A | NV 1 / NV B |
| Zero-field splitting & axial strain $D/(2\pi)$ | 2865.42 MHz | 2867.27 MHz | 2865.42 MHz | 2867.27 MHz |
| Transversal strain $E/(2\pi)$ | 0 MHz | | | |
| Transition frequencies $f_i$ (*): unused | 2932.5 MHz<br>2829.4 MHz (*) | 2751.8 MHz<br>2999.4 MHz (*) | 2571.0 MHz<br>3160.2 MHz (*) | 2990.8 MHz<br>2827.3 MHz (*) |
| Frequency separation $Min(|f_i - f_j|)$ | 66.9 MHz | | 163.5 MHz | |
| Absolute magnetic field $|B|$ | 64.23 G | 63.10 G | 105.33 G | |
| Absolute magnetic field (electron/nuclear) $|\omega_{e,i}|/(2\pi) = \gamma_e|B|/(2\pi)$ $|\omega_{n,i}|/(2\pi) = \gamma_n|B|/(2\pi)$ | 180.01 MHz /<br>19.76 kHz | 176.85 MHz /<br>19.42 kHz | 295.18 MHz /<br>32.41 kHz | |
| Misalignment $\theta_i$ | 73.42° | 45.53° | 3.58° | 74.08° |
| Quadrupole moment $Q/(2\pi)$ | -4.945 MHz [92] | | | |
| $^{14}$N Off-Axis Hyperfine coupling $A_{xx}/(2\pi) = A_{yy}/(2\pi)$ | -2.62 MHz [93] | | | |
| $^{14}$N On-Axis Hyperfine coupling $A_{zz}/(2\pi)$ | -2.162 MHz [92] | | | |
| Angle between NV centers $\beta$ | 70.53° | | | |
| Magnetic field orientation phi $\varphi$ | 47.94° | | 172.73° | |
| Effective dipole-dipole coupling strength $\nu_{dip} = g/(2\pi)$ | 0.13995 MHz | | 0.11261 MHz | |
| Computational basis of qubit system $|0\rangle, |1\rangle$ | $|g\rangle, |e_2\rangle$ | $|g\rangle, |e_1\rangle$ | $|g\rangle, |e_1\rangle$ | $|g\rangle, |e_2\rangle$ |
| Readout contrast $\alpha_i$ | 14.6 % | 14.6 % | 19.4 % | 10.7 % |
| Gyromagnetic ratio (electron/nuclear) $\gamma_e/(2\pi)$ $\gamma_n/(2\pi)$ | -28.02495 GHz/T [94]<br>3.076272 MHz/T [95] | | | |

## S.II. Simulation of Two Interacting NV Centers

### a) System Hamiltonian

We start with the Hamiltonian[80] of a single, negatively charged, NV center with an inherent nitrogen $^{14}$N nucleus in its orbital ground-state $^3\mathrm{A}_2$:

$$H^0 = \underbrace{DS_z^2 + \vec{\omega}_e \cdot \vec{S}}_{\equiv H^{\mathrm{e}}} \underbrace{+QI_z^2 + \vec{\omega}_n \cdot \vec{I}}_{\equiv H^{\mathrm{n}}} \underbrace{+\vec{S}\underline{A}\vec{I}}_{\equiv H^{\mathrm{hfc}}} \tag{S5}$$

where $D$ is the electronic zero-field splitting, $Q$ the nuclear quadrupole moment, $\underline{A}$ the hyperfine tensor, which is diagonal, $\vec{\omega}_{e/n} \equiv -\gamma_{e/n}\vec{B}$ with $\gamma_{e/n}$ being the gyromagnetic ratio of the electron/nuclear spin in angular frequency units (Hz rad/T) and $\vec{B}$ the magnetic field at the positions of the NV center. In the chosen coordinate system, the z-axis is aligned with the NV center axis. Throughout this work, we use Hamiltonians in angular frequency units, i.e. $\hbar = 1$.

We call $H^{e/\mathrm{n}}$ the electron/nuclear Hamiltonian and $H^{\mathrm{hfc}}$ the hyperfine coupling Hamiltonian. The respective spin operators $S_i(I_i)$, with $i \in \{x, y, z\}$, are defined to act on the electron (nuclear) states of the NV center only. E.g., $S_z$ is equivalent to $S_z \otimes \mathbb{1}$, i.e., a Kronecker product of $S_z$ for NV1 and the identity operator $\mathbb{1}$ for the nucleus. Both the electron spin, due to the charge state NV$^-$of the NV center, and the nuclear spin, since we consider the $^{14}$N isotope, are triplets ($S = I = 1$).

The system under consideration consists of two NV centers with distinct crystal axes (Suppl. Fig 1a sketches magnetic field setting 2 (Suppl. Table 1), in which NV 1 is more strongly misaligned), interacting with each other via dipole-dipole coupling, that are exposed to a static magnetic field and microwave pulses of arbitrary shape. Note that this section is written with respect to magnetic field setting 2. The labels NV 1 and NV 2 that indicate the order in the gate sequence, would need to be switched for magnetic field setting 1.

The free Hamiltonian of the system is given by:

$$H^{\mathrm{free}} = H_{NV1}^0 + H_{NV2}^0 + H_{12}^{int} \tag{S6}$$

with the dipole-dipole interaction Hamiltonian $H_{12}^{\mathrm{int}}$

$$H_{12}^{\mathrm{int}} = g_0(r_{12})\left[\vec{S}_1 \cdot \vec{S}_2 - 3\left(\vec{S}_1 \cdot \hat{\vec{r}}_{12}\right)\left(\vec{S}_2 \cdot \hat{\vec{r}}_{12}\right)\right] = g_0(r_{12})\vec{S}_1\underline{G}\left(\hat{\vec{r}}_{12}\right)\vec{S}_2' \tag{S7}$$

where we defined $\vec{S}_k$ as the operator $\vec{S}$ acting on the k-th NV center ($k = 1,2$) and its components $S_{i,\mathrm{k}}$ ($i = x, y, z$). The rotation of the spin operator $\vec{S}_2'$ of NV2 is detailed below.

We denote the displacement of NV center 1 from NV center 2 as $\vec{r}_{12}$ and the absolute coupling strength as $g_0(r_{12})$, depending on the distance $r_{12}$ between the NV centers:

$$g_0(r_{12}) \equiv \frac{\mu_0}{4\pi}\frac{\hbar\gamma_e^2}{r_{12}^3} \tag{S8}$$

We call $\underline{G}\left(\hat{\vec{r}}_{12}\right)$ the geometric tensor. It does not depend on the distance $r_{12}$, but on the normalized relative position vector $\hat{\vec{r}}_{12}$ (c.f. Suppl. Fig. 1a).

The two NV centers have misaligned axes, i.e. the crystal axis of NV 1 is rotated by an angle $\beta$ around the y-axis with respect to the axis NV 2 (c.f. Suppl. Fig. 1a). Resultingly, the Hamiltonian takes an unhandy shape, e.g., the zero-field splitting term reads $D(\cos(\beta)S_z + \sin(\beta)S_x)^2$. We solve this problem, by transforming the frame of NV 1: We rotate the spin operators and fields of NV 1 around the y-axis by an angle $-\beta$ into the $x', y', z'$-frame where the axis of NV 1 coincides with the $z'$-axis (c.f. Suppl. Fig. 1a):

$$\vec{\omega}' = R_y(-\beta)\,\vec{\omega}, \qquad \vec{S}' = R_y(-\beta)\,\vec{S}, \qquad \vec{S} \equiv \begin{pmatrix} S_x \\ S_y \\ S_z \end{pmatrix}, \qquad R_y(-\beta) = \begin{pmatrix} \cos\beta & 0 & -\sin\beta \\ 0 & 1 & 0 \\ \sin\beta & 0 & \cos\beta \end{pmatrix} \tag{S9}$$

In the primed frame, the Hamiltonian of NV 1 takes the again the usual form as given in Suppl. Equation S5, i.e. the zero-field splitting term reads $D(S_z')^2$. In this picture the NV centers experience a different magnetic field and a modified geometric tensor $\underline{G}(\hat{\vec{r}}_{12})$.

As the NV centers are exposed to a constant magnetic field that is (generally) misaligned with their crystal axes, their quantization axes are changed (in length and direction), which results in new eigenvalues and eigenstates ($|g\rangle, |e_1\rangle, |e_2\rangle$) of the Hamiltonian. As a consequence, the eigenstates of the electron Hamiltonian $H^e$ do not coincide with the eigenstates of the $S_{z,2}, S_{z,1}'$ operators, but we can define new spin operators $\tilde{S}_{z,2}, \tilde{S}_{z,1}'$ that have the same eigenstates as $H^{\mathrm{e}}$ and take the usual (matrix) form in the corresponding basis. Similarly, we define $\tilde{S}_{x/y,2}, \tilde{S}_{x/y,1}'$.

The dipole-dipole interaction is (typically) small compared to the electron transition frequencies ($g_0 \ll \omega_{\mathrm{i}}$), so that a secular approximation can be applied, whereby small, non-diagonal elements (in the basis of the $\tilde{S}_{z,i}$ eigenstates) in the dipole-dipole interaction Hamiltonian are neglected. This simplifies the dipole-dipole coupling to:

$$H_{12}^{\mathrm{int}} = g(\vec{r}_{12}, \beta, \vec{B})\,\tilde{S}_{z,2}\tilde{S}_{z,1}' \tag{S10}$$

where $g(\vec{r}_{12}, \beta, \vec{B})$ is the effective coupling strength ($\nu_{dip} = g/(2\pi)$ in frequency units).

By transforming the frame of NV center 1, transforming the spin operators into the basis of the electron eigenstates and performing the secular approximation, we have developed an effective picture of the two NV center system (c.f. Suppl. Fig. 1b).

The total (noiseless) Hamiltonian is given by:

$$H_{driven}(t) = H^{\mathrm{free}} + H^{\mathrm{mw}}(t) \tag{S11}$$

where $H_{NV1}^0$ and $H_{NV2}^0$ are the single NV center Hamiltonians of NV centers 1 and 2, and $H^{\mathrm{mw}}$ the microwave Hamiltonian

$$H^{\mathrm{mw}}(t) = \sum_{i=1,2} \sqrt{2}\left[\Omega_{i,x}(t)\cos(\omega_i t + \xi_i) + \Omega_{i,y}(t)\sin(\omega_i t + \xi_i)\right] H^{\mathrm{control}} \tag{S12}$$

where $\Omega_{i,x}(t), \Omega_{i,y}(t)$ are time-dependent amplitudes (or control functions), that vary only slowly compared to the angular carrier frequencies $\omega_i$. Phases of the carrier, generally set to zero, are denoted $\xi_i$ and $H^{\text{control}}$ is the control Hamiltonian

$$H^{\text{control}} \equiv \tilde{S}'_{x,1} + \tilde{S}_{x,2} + \tilde{\gamma}(I_{x,1} + I_{x,2}) \tag{S13}$$

with $\tilde{\gamma} \equiv \gamma_n/\gamma_e = \omega_n/\omega_e$.

### b) Simulation

For simulating the time dynamics of the system, we start with single NV Hamiltonians $H^0_{1/2}$, with the rotation by angle $-\beta$ applied on NV center 1, as above and compute the matrix $T$ that diagonalizes the sum of the electron Hamiltonians $H^e \equiv H^e_1 + H^e_2$, such that $T^\dagger H^e T$ is diagonal.

We apply the diagonalization matrix to the sum of the single NV Hamiltonians $H^0_1 + H^0_2$ and then add the effective dipole-dipole coupling term in Suppl. Equation S8 to obtain the (effective) free Hamiltonian,

$$H_{\text{free}} = T^\dagger(H^0_1 + H^0_2)T + H^{\text{int}}_{12} \tag{S14}$$

From here on, we use the basis of the electronic eigenstates, so that $H^{\text{int}}_{12} = g\, \tilde{S}_{z,2}\tilde{S}'_{z,1}$ as above.

The spin operators are defined as

$$\begin{aligned} \tilde{S}_{i,2} &= S_i \otimes \mathbb{1} \otimes \mathbb{1} \otimes \mathbb{1}, \qquad \tilde{S}'_{i,1} = \mathbb{1} \otimes \mathbb{1} \otimes S_i \otimes \mathbb{1} \\ I_{i,2} &= \mathbb{1} \otimes I_i \otimes \mathbb{1} \otimes \mathbb{1}, \qquad I_{i,1} = \mathbb{1} \otimes \mathbb{1} \otimes \mathbb{1} \otimes I_i \end{aligned} \tag{S15}$$

With order (NV2, nuclear spin 2, NV1, nuclear spin 1) and

$$S_x = \frac{1}{\sqrt{2}}\begin{pmatrix} 0 & 1 & 0 \\ 1 & 0 & 1 \\ 0 & 1 & 0 \end{pmatrix} = I_x, \qquad S_y = \frac{i}{\sqrt{2}}\begin{pmatrix} 0 & -1 & 0 \\ 1 & 0 & -1 \\ 0 & 1 & 0 \end{pmatrix} = I_y, \qquad S_z = \begin{pmatrix} 1 & 0 & 0 \\ 0 & 0 & 0 \\ 0 & 0 & -1 \end{pmatrix} = I_z \tag{S16}$$

If the magnetic field is static, the Hamiltonian of the system is given by $H_{\text{free}}$ and therefore time-independent. In this case, the time evolution operator is calculated via the matrix exponential $U(t) = \exp(-iH_{\text{free}}\, t)$ and then transformed into the rotating frame via:

$$U_{\text{rot}}(t) = V^\dagger(t)U(t) \quad \text{with} \quad V(t) = \exp(-iH_{\text{trans}}\, t) \tag{S17}$$

where $H_{\text{trans}} \equiv \omega_1 \left(\tilde{S}'_{z,1}\right)^2 + \omega_2\left(\tilde{S}_{z,2}\right)^2$ is the Hamiltonian that generates the rotating frame transformation, $\omega_{1/2} = 2\pi f_{1/2}$ are the angular carrier frequencies, ideally coinciding with the transition lines $2\pi f_{\text{i}}$ of the chosen qubit system (Suppl. Table 1).

If microwave pulses are applied, the Hamiltonian in Suppl. Equation S11 contains time-dependent terms. In order to calculate the time evolution from $t_0$ to $t$, we approximate the time-evolution via the Riemann summation method, whereby we evaluate the Hamiltonian at $N$ discrete time points:

$$
\begin{aligned}
U(t, t_0) &= \mathcal{T}\exp\left(-\mathrm{i}\int_{t_0}^{t} H_{\mathrm{rot}}(t')\mathrm{d}t'\right) \\
&\approx e^{-iH_{\mathrm{rot}}(t_N)\tau} \ldots e^{-iH_{\mathrm{rot}}(t_2)\tau} e^{-iH_{\mathrm{rot}}(t_1)\tau}
\end{aligned}
\tag{S18}
$$

with $t_j = t_0 + j\,\tau$ and $\tau = (t - t_0)/N$, where $N = 100$ GHz $(t - t_0)$ and $\mathcal{T}$ is the time ordering operator.

To realize faster convergence, i.e. realize the same precision with larger time-steps, the Hamiltonian is transformed into the rotating frame (at every timestep) in which the Hamiltonian varies only slowly:

$$
H_{\mathrm{rot}}(t_i) = V_i^{\dagger}(H_{\mathrm{driven}}(t_i) - H_{\mathrm{trans}})V_i \quad \text{with} \quad V_i = V(t_i) \tag{S19}
$$

When dealing with sequences of pulses and free evolution, we multiply the respective unitaries time-orderly.

Readout results $R(\rho)$ are obtained via the POVM element $E_g(\{\alpha_i\})$, similar to the definition in [21], as:

$$
\begin{aligned}
R(\rho) &= \mathrm{Tr}_e\left[E_g \mathrm{Tr}_n\left(U\rho U^{\dagger}\right)\right] - \mathrm{Tr}_e\left[E_g \mathrm{Tr}_n\left(X_{12}U\rho U^{\dagger}X_{12}^{\dagger}\right)\right] \\
E_g(\{\alpha_i\}) &= \frac{1}{\alpha_1 + \alpha_2}\left(\alpha_1 |g\rangle\langle g| \otimes \mathbb{1} + \alpha_2 \mathbb{1} \otimes |g\rangle\langle g|\right)
\end{aligned}
\tag{S20}
$$

where $X_{12}$ is a $\pi_x$-pulse on NV center 1, followed by a $\pi_x$-pulse on NV center 2, $\mathrm{Tr}_{e/n}$ is the trace of the electron/nuclear spins, and $\alpha_{1/2}$ is a contrast parameter, proportional to the spin contrast observed in experiment.

As we only read out diagonal elements in the measurement, the results are the same in the rotating frame as in the lab frame (since $V_i$ is diagonal).

Within the simulation, if not stated otherwise, the electron state is perfectly initialized in the ground state, the nuclear state in the maximally mixed state (thermal state approximation). If the nuclear state is stated to be initialized, then it is initialized into its $m_I = 0$ state.

If not stated otherwise, the pulses are chosen to have a sinusoidal shape. E.g., for a $\pi_x$-pulse on NV center 1, the control functions are given by:

$$
\Omega_{1,x}(t) = \Omega_{max} sin(\pi \cdot t/t_{\pi}),\ \Omega_{1,y}(t) = \Omega_{2,x}(t) = \Omega_{2,y}(t) = 0 \tag{S21}
$$

where $\Omega_{max} = \frac{\pi}{2}\Omega_{rabi}$ is the peak amplitude of the envelope, $\Omega_{rabi} = \frac{\pi}{t_{\pi}}$, and $t_{\pi}$ is the $\pi$ pulse duration of the sine shaped pulse.

For evaluations of the gate fidelity, we use the definition of the average fidelity, given in [96]:

$$F_{\mathrm{avg}} = \frac{1}{d(d+1)} \sum_{i,j=1}^{d} \left( \langle i | G^{\dagger} \mathcal{D}(|i\rangle\langle j|) G | j \rangle + \langle i | G^{\dagger} \mathcal{D}(|j\rangle\langle j|) G | i \rangle \right) \tag{S22}$$

where $G$ is the target quantum gate and $\mathcal{D}$ is the dynamical map, i.e. the time evolution of the electron spins (under the influence of the nuclear spins):

$$\mathcal{D}(\rho) = \mathrm{Tr}_n \left( U \rho U^{\dagger} \right) \tag{S23}$$

In our case, $d = 4$ and the $\{|i\rangle\}$ states are the basis of the qubit subspace. We evaluate the gate-fidelity in the rotating frame.

The randomized benchmarking circuits, used in the simulation, are the same as in the corresponding experiments. When sweeping the Rabi frequency, the set of pulses and the free evolutions are calculated for each Rabi frequency individually. When crosstalk and leakage are stated to be turned off, this means that the control Hamiltonian is modified to only act on the qubit levels of the desired NV center.

## S.III. Single-Qubit Randomized Benchmarking

To estimate the entangling gate fidelity $EPG_{2q}$ from the measured EPC of the gate set, we need to separate the error contribution of the single-qubit gates, as described in the Methods section:

$$EPC = 1 - \left(1 - EPG_{2q}\right)^{GPC_{2q}} \left(1 - EPG_{1q}\right)^{GPC_{1q}} \tag{S24}$$

This is often done by measuring single-qubit randomized benchmarking on the sub systems (Suppl. Fig. 2a). Due to a particularity of our system, we take a slightly different approach here. We first note that due to limitations of our microwave hardware, we never apply pulses on both NVs at the same time. As a result of the unpolarized nitrogen spins, a single qubit rotation on NV1 thus can cause an unwanted phase pickup during the free evolution on NV2 and vice versa. This effect is neglected when carrying our single-qubit randomized benchmarking on a single NV only.

To obtain a one-qubit fidelity that represents all errors that are present when using the two-qubit register, we determine an average effective single-qubit error $EPC_{1q} = 1 - \left(1 - EPG_{1q}\right)^{GPC_{1q}}$ from a modified two-qubit randomized benchmarking experiment in Suppl. Fig. 2b: This approach excludes all entangling gates from the generated experiment ($GPC_{2q} = 0$) and appends single-qubit correction pulses to revert to the ground state (resulting modified $GPC_{1q} = 11.5$). On every qubit, the modified experiment retains a random sequence of single-qubit Clifford operations – except for the free evolution that we treat as an error. We then extract the average effective single-qubit error $\mathrm{EPC_{1q}}$ from the fitted decay lifetime parameter $p = \exp\left(-\frac{1}{N_d}\right)$ as:

$$EPC_{1q} = (2^{\mathrm{n}} - 1)/2^{n}(1 - p) = 0.085 \pm 0.013 \tag{S25}$$

With n the number of qubits in the Clifford sequence, here $n = 2$. This translates to an effective single-qubit gate fidelity of $F_{1q}$=(99.23 $\pm$ 0.12) %.

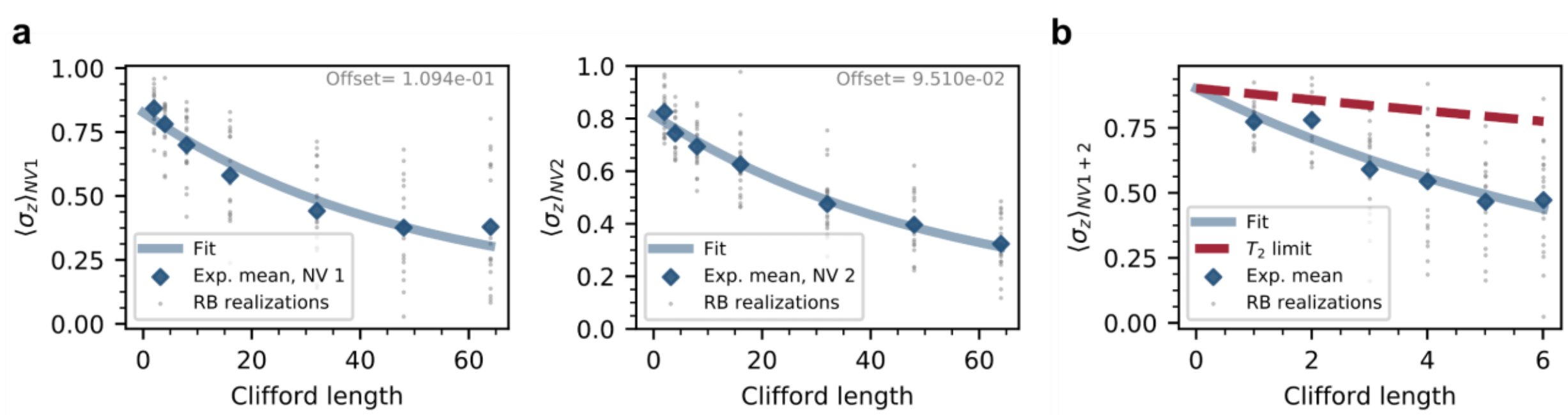

Suppl. Figure 2. Single-qubit randomized benchmarking. (a) Given the gate set $\mathrm{GPC_{1q}} = 1.97$, we extract bare single-qubit gate fidelities of $F_{1q,NV1} = (99.49 \pm 0.56)\%$ and $F_{1q,NV2} = (99.53 \pm 0.49)\%$. Offset parameters are fixed by a separate measurement, as described in the Methods section. (b) Average single-qubit error $F_{1q} = (99.23 \pm 0.12)\%$ from modified two-qubit randomized benchmarking with stripped entangling gate.

## S.IV. Entangling Gate Dynamics & Calibration

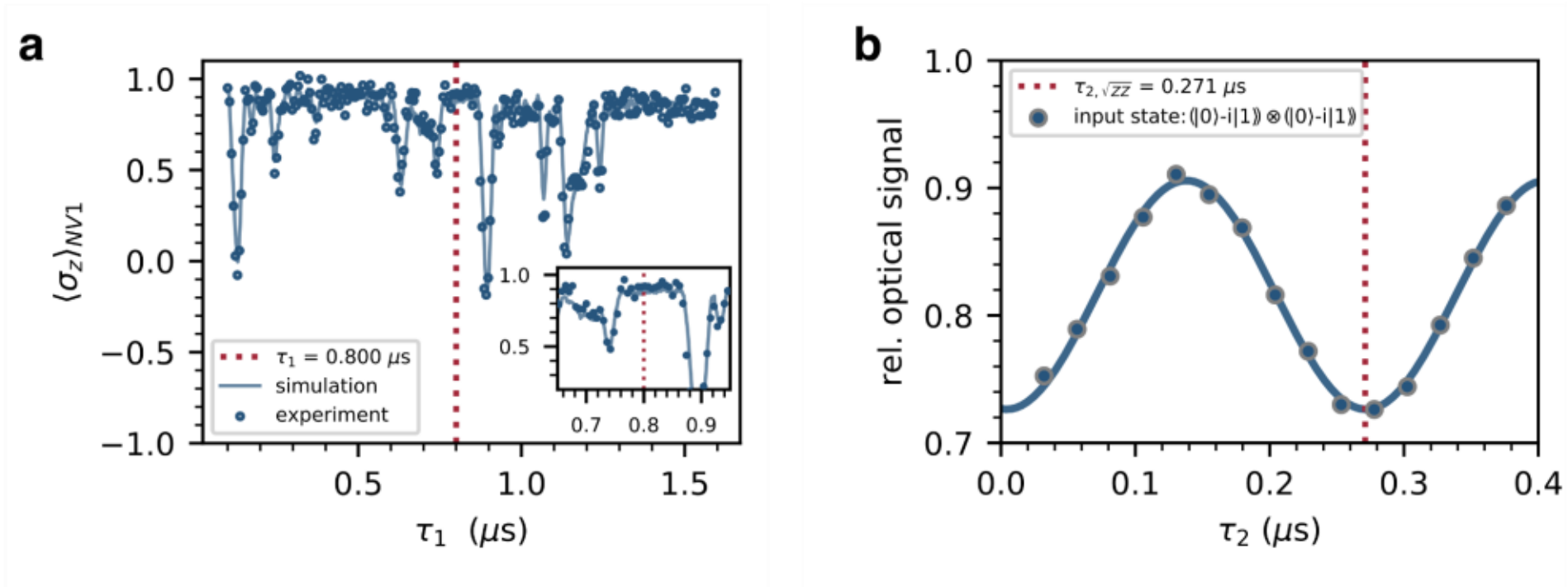

Suppl. Figure 3. Entangling gate calibration. (a) XY8-1 experiment performed with pulse spacing $\tau_1$ on the misaligned target NV1 at magnetic field settings 2. (b) Measured repeated application of the $\sqrt{ZZ}$ gate with varying $\tau_2$ ($\tau_1$=800 ns, $n_{rep} = 4$, $\Omega_{rabi}/(2\pi) = 22.97$ MHz) for calibration. The projection pulse after the $\sqrt{ZZ}$ gates for both NVs is a $\pi_x/2$. The solid line is a fit from which we extract the $\tau_{2,\sqrt{ZZ}}$ that realizes our entangling gate.

In order to calibrate the duration of our entangling gate, we first analyze how it scales with the duration of our decoupling sequence and the choice of $\tau_1$ and $\tau_2$.

To this end, we simplify the effective model Hamiltonian (Suppl. Equations S6, S11) further by treating the effect of the nitrogen nuclear spin as effective detuning (similarly to [97]) and arrive at:

$$\begin{gathered} H = H^0_{NV1} + H^0_{NV2} + H^{\mathrm{mw}} + H^{\mathrm{int}}_{12} \\ H^0_{NV1} = \mathrm{D}S^2_{z,1} + \Delta_1 S_{z,1} \\ H^0_{NV2} = \mathrm{D}S^2_{z,2} + \Delta_2 S_{z,2} \\ H^{\mathrm{mw}} = \sqrt{2}\,\{\Omega_{1,x}(t)\cos(\omega_1 t + \xi_1) + \Omega_{2,x}(t)\cos(\omega_2 t + \xi_2)\}(S_{x,1} + S_{x,2}) \\ H^{\mathrm{int}}_{12} = g S_{z,2} S_{z,1} \end{gathered} \tag{S26}$$

Here, we are considering the effective dipolar coupling $g = 2\pi\nu_{dip}$ between the NV centers[98], with the zero-field splitting D of the NVs, $\Delta_k, k = 1,2$ is the effective detuning due to the Zeeman and hyperfine splittings, whose effect is typically refocused after the gate, $\Omega_{1,x}(t)$ and $\Omega_{2,x}(t)$ characterize the Rabi frequencies of the microwave fields with angular frequencies $\omega_1$ and $\omega_2$ and initial relative phases $\xi_1$ and $\xi_2$ that we use to drive the two NV centers during the refocusing pulses. We use here the operator labels $S_{z,1}$ and $S_{z,2}$ for $\tilde{S}'_{z,1}$ and $\tilde{S}_{z,2}$ from section 2b, respectively, for simplicity of presentation and without loss of generality.

We move to the rotating frame, defined by $\omega_1 S^2_{z,1}$ and $\omega_2 S^2_{z,2}$, apply the rotating wave-approximation, using that $\Omega_{1,2}(t) \ll \omega_{1,2}$, and obtain

$$H_{rot}^{NV1} = (\mathrm{D} - \omega_1)S_{z,1}^2 + \Delta_1 S_{z,1} + \frac{\Omega_1(t)}{\sqrt{2}} S_{x,1} = \begin{pmatrix} \mathrm{D} - \omega_1 + \Delta_1 & \frac{\Omega_1(t)}{2} & 0 \\ \frac{\Omega_1(t)}{2} & 0 & \frac{\Omega_1(t)}{2} \\ 0 & \frac{\Omega_1(t)}{2} & \mathrm{D} - \omega_1 - \Delta_1 \end{pmatrix}$$

$$= \begin{pmatrix} \delta_1 & \frac{\Omega_1(t)}{2} & 0 \\ \frac{\Omega_1(t)}{2} & 0 & \frac{\Omega_1(t)}{2} \\ 0 & \frac{\Omega_1(t)}{2} & \delta\delta_1 \end{pmatrix} \tag{S27}$$

$$H_{rot}^{NV2} = (\mathrm{D} - \omega_2)S_{z,2}^2 + \Delta_2 S_{z,2} + \frac{\Omega_2(t)}{\sqrt{2}} S_{x,2} = \begin{pmatrix} \mathrm{D} - \omega_2 + \Delta_2 & \frac{\Omega_2(t)}{2} & 0 \\ \frac{\Omega_2(t)}{2} & 0 & \frac{\Omega_2(t)}{2} \\ 0 & \frac{\Omega_2(t)}{2} & \mathrm{D} - \omega_2 - \Delta_2 \end{pmatrix}$$

$$= \begin{pmatrix} \delta\delta_2 & \frac{\Omega_2(t)}{2} & 0 \\ \frac{\Omega_2(t)}{2} & 0 & \frac{\Omega_2(t)}{2} \\ 0 & \frac{\Omega_2(t)}{2} & \delta_2 \end{pmatrix} \tag{S28}$$

$$H_{rot}^{total} = H_{rot}^{NV1} + H_{rot}^{NV2} + g\, S_{z,2}S_{z,1}$$

$$= \begin{pmatrix} \delta_1 + \delta\delta_2 + g & \frac{\Omega_1(t)}{2} & 0 & \frac{\Omega_2(t)}{2} & 0 & 0 & 0 & 0 & 0 \\ \frac{\Omega_1(t)}{2} & \delta\delta_2 & \frac{\Omega_1(t)}{2} & 0 & \frac{\Omega_2(t)}{2} & 0 & 0 & 0 & 0 \\ 0 & \frac{\Omega_1(t)}{2} & \delta\delta_1 + \delta\delta_2 - g & 0 & 0 & \frac{\Omega_2(t)}{2} & 0 & 0 & 0 \\ \frac{\Omega_2(t)}{2} & 0 & 0 & \delta_1 & \frac{\Omega_1(t)}{2} & 0 & \frac{\Omega_2(t)}{2} & 0 & 0 \\ 0 & \frac{\Omega_2(t)}{2} & 0 & \frac{\Omega_1(t)}{2} & 0 & \frac{\Omega_1(t)}{2} & 0 & \frac{\Omega_2(t)}{2} & 0 \\ 0 & 0 & \frac{\Omega_2(t)}{2} & 0 & \frac{\Omega_1(t)}{2} & \delta\delta_1 & 0 & 0 & \frac{\Omega_2(t)}{2} \\ 0 & 0 & 0 & \frac{\Omega_2(t)}{2} & 0 & 0 & \delta_1 + \delta_2 - g & \frac{\Omega_1(t)}{2} & 0 \\ 0 & 0 & 0 & 0 & \frac{\Omega_2(t)}{2} & 0 & \frac{\Omega_1(t)}{2} & \delta_2 & \frac{\Omega_1(t)}{2} \\ 0 & 0 & 0 & 0 & 0 & \frac{\Omega_2(t)}{2} & 0 & \frac{\Omega_1(t)}{2} & \delta_2 + \delta\delta_1 + g \end{pmatrix} \tag{S29}$$

where we assume that the driving field frequency $\omega_1(\omega_2)$ is highly detuned from the transition frequencies of NV2 (NV1), so there is negligible crosstalk from pulses applied to NV1 on NV2 and vice versa. In addition, we take $\xi_1 = \xi_2 = 0$ for simplicity of presentation and without loss of generality. The matrix representation of $H_{rot}^{NV1}, H_{rot}^{NV2}$ is in the single qutrit basis of each of the NV centers, while $H_{rot}^{total}$ is represented in the two-qutrit basis. In the experiment, the first driving field is approximately resonant with the $|0\rangle \leftrightarrow |+1\rangle$ transition on NV1 ($\delta_1 \equiv \mathrm{D} - \omega_1 + \Delta_1 \ll \Omega_1(t), \delta\delta_1 \equiv \mathrm{D} - \omega_1 - \Delta_1 \gg \Omega_1(t)$) and the second field - with the $|0\rangle \leftrightarrow |-1\rangle$ transition on NV2 ($\delta_2 \equiv \mathrm{D} - \omega_2 - \Delta_2 \ll \Omega_2(t), \delta\delta_2 \equiv \mathrm{D} - \omega_2 + \Delta_2 \gg \Omega_2(t)$). The Hamiltonian elements, which characterize the evolution of the states, which we can effectively address with the microwave fields (they do not have a very high frequency offset in the diagonal element of the Hamiltonian) are highlighted with dashed lines. Thus, the respective, reduced Hamiltonian, which characterizes the evolution that leads to our two-qubit gate is given by

$$H_{rot,reduced}^{total} = \begin{pmatrix} \delta_1 & \frac{\Omega_1(t)}{2} & \frac{\Omega_2(t)}{2} & 0 \\ \frac{\Omega_1(t)}{2} & 0 & 0 & \frac{\Omega_2(t)}{2} \\ \frac{\Omega_2(t)}{2} & 0 & \delta_1 + \delta_2 - g & \frac{\Omega_1(t)}{2} \\ 0 & \frac{\Omega_2(t)}{2} & \frac{\Omega_1(t)}{2} & \delta_2 \end{pmatrix} \tag{S30}$$

The detunings $\delta_1$ and $\delta_2$ typically cause dephasing but can be refocused by applying dynamical decoupling. The z-z coupling $g$ allows us to apply the $\sqrt{ZZ}$ gate, as evident from the analysis below. The Hamiltonian during free evolution without microwave pulses and the respective propagator are given by

$$H_{free} = \begin{pmatrix} \delta_1 & 0 & 0 & 0 \\ 0 & 0 & 0 & 0 \\ 0 & 0 & \delta_1 + \delta_2 - g & 0 \\ 0 & 0 & 0 & \delta_2 \end{pmatrix}, \; U_{free}(T_f) = \exp(-i\, H_{free} T_f) = \begin{pmatrix} e^{-iT_f\delta_1} & 0 & 0 & 0 \\ 0 & 1 & 0 & 0 \\ 0 & 0 & e^{iT_f(-\delta_1 - \delta_2 + g)} & 0 \\ 0 & 0 & 0 & e^{-i\delta_2 T_f} \end{pmatrix}, \tag{S31}$$

where $T_f$ is the respective free evolution time. We assume that the Rabi frequencies are much stronger than $g, \delta_1$ and $\delta_2$, so we can neglect their effect during the refocusing pulses. Then, the Hamiltonians and propagators of the π pulses on NV 1 and NV 2 are respectively

$$H_{pulse}^{NV1} \approx \begin{pmatrix} 0 & \frac{\Omega_1(t)}{2} & 0 & 0 \\ \frac{\Omega_1(t)}{2} & 0 & 0 & 0 \\ 0 & 0 & 0 & \frac{\Omega_1(t)}{2} \\ 0 & 0 & \frac{\Omega_1(t)}{2} & 0 \end{pmatrix}, \; U_1(\theta_1) = \exp\left(-i\, H_{pulse}^{NV1} T_1\right) \approx \begin{pmatrix} \cos\left(\frac{\theta_1}{2}\right) & -i\sin\left(\frac{\theta_1}{2}\right) & 0 & 0 \\ -i\sin\left(\frac{\theta_1}{2}\right) & \cos\left(\frac{\theta_1}{2}\right) & 0 & 0 \\ 0 & 0 & \cos\left(\frac{\theta_1}{2}\right) & -i\sin\left(\frac{\theta_1}{2}\right) \\ 0 & 0 & -i\sin\left(\frac{\theta_1}{2}\right) & \cos\left(\frac{\theta_1}{2}\right) \end{pmatrix} \tag{S32}$$

$$H_{pulse}^{NV2} \approx \begin{pmatrix} 0 & 0 & \frac{\Omega_2(t)}{2} & 0 \\ 0 & 0 & 0 & \frac{\Omega_2(t)}{2} \\ \frac{\Omega_2(t)}{2} & 0 & 0 & 0 \\ 0 & \frac{\Omega_2(t)}{2} & 0 & 0 \end{pmatrix}, \; U_2(\theta_2) = \exp\left(-i\, H_{pulse}^{NV2} T_2\right) \approx \begin{pmatrix} \cos\left(\frac{\theta_2}{2}\right) & 0 & -i\sin\left(\frac{\theta_2}{2}\right) & 0 \\ 0 & \cos\left(\frac{\theta_2}{2}\right) & 0 & -i\sin\left(\frac{\theta_2}{2}\right) \\ -i\sin\left(\frac{\theta_2}{2}\right) & 0 & \cos\left(\frac{\theta_2}{2}\right) & 0 \\ 0 & -i\sin\left(\frac{\theta_2}{2}\right) & 0 & \cos\left(\frac{\theta_2}{2}\right) \end{pmatrix} \tag{S33}$$

where $\theta_1 = \Omega_1 T_1 = \pi$ and $\theta_2 = \Omega_2 T_2 = \pi$ (we assumed that the pulses are rectangular for simplicity of presentation and without loss of generality).

In order to characterize the gate, we consider a sequence of two instantaneous $\pi$ pulses on each NV and analyze the evolution in the interaction basis of the pulses. The Hamiltonians in each free evolution period are given by:

Period 1: before the first π pulse on NV1:

$$H_{int,1} = H_{free} = \begin{pmatrix} \delta_1 & 0 & 0 & 0 \\ 0 & 0 & 0 & 0 \\ 0 & 0 & -g+\delta_1+\delta_2 & 0 \\ 0 & 0 & 0 & \delta_2 \end{pmatrix}, t_1 = \frac{\tau_1}{2}, \tag{S34}$$

Period 2: after one π pulse on NV1, no π pulses on NV2:

$$H_{int,2} = U_{int,2}\, H_{free}\, U_{int,2}^{\dagger} = \begin{pmatrix} 0 & 0 & 0 & 0 \\ 0 & \delta_1 & 0 & 0 \\ 0 & 0 & \delta_2 & 0 \\ 0 & 0 & 0 & -g+\delta_1+\delta_2 \end{pmatrix}, \text{where } U_{int,2} = U_1(\pi), t_2 = \frac{\tau_1}{2} - \tau_2 \tag{S35}$$

Period 3: after one π pulse on NV1, one π pulse on NV2:

$$H_{int,3} = U_{int,3}\, H_{free}\, U_{int,3}^{\dagger} = \begin{pmatrix} \delta_2 & 0 & 0 & 0 \\ 0 & -g+\delta_1+\delta_2 & 0 & 0 \\ 0 & 0 & 0 & 0 \\ 0 & 0 & 0 & \delta_1 \end{pmatrix}, \text{where } U_{int,3} = U_2(\pi)U_1(\pi), t_3 = \frac{\tau_1}{2} + \tau_2 \tag{S36}$$

Period 4: after two π pulses on NV1, one π pulse on NV2:

$$H_{int,4} = U_{int,4}\, H_{free}\, U_{int,4}^{\dagger} = \begin{pmatrix} -g+\delta_1+\delta_2 & 0 & 0 & 0 \\ 0 & \delta_2 & 0 & 0 \\ 0 & 0 & \delta_1 & 0 \\ 0 & 0 & 0 & 0 \end{pmatrix}, \text{where } U_{int,4} = U_1(\pi)\, U_2(\pi)U_1(\pi), t_4 = \frac{\tau_1}{2} - \tau_2 \tag{S37}$$

Period 5: after two π pulses on NV1, two π pulses on NV2:

$$H_{int,5} = U_{int,5}\, H_{free}\, U_{int,5}^{\dagger} = \begin{pmatrix} \delta_1 & 0 & 0 & 0 \\ 0 & 0 & 0 & 0 \\ 0 & 0 & -g+\delta_1+\delta_2 & 0 \\ 0 & 0 & 0 & \delta_2 \end{pmatrix}, \text{where } U_{int,5} = U_2(\pi)U_1(\pi)\, U_2(\pi)U_1(\pi), t_5 = \tau_2 \tag{S38}$$

The total evolution in the interaction basis is then given by

$$U_{int} = \exp\left(-i\sum H_{int,k} t_k\right) = \exp(\mathrm{i}\,\xi_{\mathrm{g}}) \begin{pmatrix} 1 & 0 & 0 & 0 \\ 0 & e^{iN_\pi g \tau_2} & 0 & 0 \\ 0 & 0 & e^{iN_\pi g \tau_2} & 0 \\ 0 & 0 & 0 & 1 \end{pmatrix} \tag{S39}$$

where we used that $[H_{int,i}, H_{int,j}] = 0$ to obtain the first equality, $\xi_{\mathrm{g}} = N_\pi[\frac{g}{2}\left(-\frac{\tau_1}{2} + \tau_2\right) + \frac{\tau_1}{2}(\delta_1 + \delta_2)]$ is an irrelevant global phase and $N_\pi$ is the total number of applied pulses on each NV center, e.g., $N_\pi = 2$ in our example with two pulses or $N_\pi = 8$ for XY8-1 and $N_\pi = 16$ for XY8-2. Thus, in the following, we define the effective evolution time $t_{evol} = N_\pi \tau_2$.

For demonstrating the controlled oscillations of the $\sqrt{ZZ}$ gate with XY8-1 decoupling on the Bloch sphere equator in Fig. 2, we use $\tau_1 = 3000$ ns. This choice allows to sweep the effective evolution time $t_{evol} = N_\pi \tau_2$ with large dynamic range (constrained by $-\tau_1/2 \leq \tau_2 \leq \tau_1/2$), but is not an optimal choice in terms of fidelity for two reasons:

First, shorter $\tau_1$ allow to realize the same gate unitary with shorter total sequence length ($= N_\pi \tau_1$, independent of $\tau_2$), and thus less decoherence. Consequently, for realizing the $\sqrt{ZZ}$ unitary, $\tau_1$ should be chosen close to $\tau_{1,min} = \frac{2}{4 N_\pi v_{dip}} \sim 520$ ns for our $v_{dip} = 120$ kHz, $N_\pi = 8$ when using XY8-1. We note that using shorter $\tau_1$ for our entangling gate is also possible by increasing $N_\pi$ in the dynamical decoupling sequence, e. g., using XY8-2 that consists of 16 pulses, as long as $t_{evol} = N_\pi \tau_2$ with $\tau_2 \leq \tau_1/2$ is kept constant. Decreasing the pulse spacing $\tau_1$ should improve fidelity as the characteristic $T_2$ time of a decoupling sequence typically increases when the pulse separation is shorter[99–101].

Second, due to the misaligned magnetic field on the target qubit, population transfer to and entanglement with the nitrogen nuclear spin can occur. Since we can choose $\tau_1$ freely (but not $\tau_2$ that is set by $v_{dip}$), we use this degree of freedom to minimize such an unwanted interaction with the nitrogen spin. In Suppl. Fig. 3a, we probe the polarization loss due to interaction with the nitrogen by initializing the target qubit into superposition and applying a standard XY8-1 experiment only on the target qubit with varying $\pi$ pulse spacing $\tau_1$. We observe that the simulation is well reproducing the experimental data and find a $\tau_1 = 800$ ns that reduces interaction with the nitrogen spin. In future, modified gate microwave sequence could minimize this interaction with the nucleus more strictly[102].

After determining $\tau_1$, our gate calibration for $\tau_2$ is carried out as follows: We initialize the input state ($|0\rangle - i|1\rangle \otimes |0\rangle - i|1\rangle$) and apply the $\sqrt{ZZ}$ gate sequence, with varying $\tau_2$, four times ($n = 4$). Repeating the gate improves the accuracy of the calibration, as more oscillation periods are recorded for a $\tau_2$ sweep. To the experimentally recorded oscillation (eg. in Suppl. Fig. 3b), we fit a sine and find the $\tau_{2,\sqrt{ZZ}}$ that is realizing minimal fluorescence, as expected from a quarter rotation (repeated $n = 4$ times) on the Bloch sphere equator.

In order to capture the gate dynamic accurately, we repeat the same calibration procedure to obtain a coupling parameter for our model. In addition to the experiment in in Suppl. Fig. 3b at magnetic field setting 2, we experimentally perform a calibration also at magnetic field setting 1 ($\tau_1$=800 ns, $n = 4$, $\Omega_{rabi}/(2\pi) = 15.51$ MHz, data not shown) and obtain a calibrated $\tau_{2,\sqrt{ZZ},B_1} = 212.6$ ns. We then simulate both calibration experiments individually and find the effective dipolar coupling parameters $v_{dip}$ in Suppl. Table 1 that reproduce the experimentally observed $\tau_{2,\sqrt{ZZ},B_1}, \tau_{2,\sqrt{ZZ},B_2}$.

The dipole-dipole coupling of 119.8 kHz sets an upper limit of $(9.54 \pm 0.03)$ nm on the distance between the NV centers[103]. An exact distance estimate would require knowledge the relative orientation of the dipoles and could be obtained by repeated DEER measurements in a varying bias magnetic field[104].

## S.V. Charge Initialization & Readout

Our charge initialization is based on a weak orange laser pulse and post-selection of the recorded spin readout photons. After the green (552 nm) spin initialization laser pulse, the charge state of a single NV in bulk diamond reaches a steady state occupation of typically $A(NV^-)/(A(NV^0)+A(NV^-)) \sim 0.7$ [38], where $A$ denotes the probability to find the NV in a certain charge state. The orange laser probes the $NV^-$ absorption with minimal overlap to the $NV^0$ absorption spectrum. At the same time, its weak power avoids ionization. Thus, more readout photons are expected during the orange laser pulse if both of the NVs are in the negative charge state. In Suppl. Fig. 4a, we measure readout photons per 3.5 ms long orange laser on our coupled NV quantum register after spin initialization with a 3 µs green laser. Due to the limited collection efficiency, the three expected Poissonian peaks $\{(NV^0, NV^0);\ (NV^0, NV^-);\ (NV^-, NV^-)\}$ overlap in the histogram. For determination of the charge state, we fit a Poissonian model to the photon probability $p(n_{phot})$ with free weighting parameters $A_i$ and fitted mean rates $\lambda_i$:

$$p(n_{phot}) = A_{--}P(\lambda_{--}) + A_{-0}P(\lambda_{-0}) + A_{00}P(\lambda_{00}) \tag{S40}$$

The extracted weights of the charge state slightly differ from the values $A_{00} = 0.09,\ A_{-0} = 0.42, A_{--} = 0.49$ expected for an independent combination of two bulk NVs. We speculate that a single charge shared between both NVs is more favorable in our system.

In Suppl. Fig. 4b we append another 2.9 ms long orange laser pulse to independently probe the charge state fidelity after the $t_{pcs} = 3.5$ ms long charge initialization laser pulse, without applying post selection yet. As expected, we recover charge state weights that are in agreement with the data in Suppl. Fig. 4a.

We then continue in Suppl. Fig. 4c to analyze only charge readout data for experimental shots where the counted photons during the first initialization orange laser pulse satisfy a threshold photon $n_{phot} > n_{thresh}$. After applying post-selection with the experimental settings ($t_{pcs} = 3.5$ ms, $n_{thresh}$=9) presented in Fig. 2,3, we measure an increase in $(NV^-, NV^-)$ charge state initialization fidelity from $(39 \pm 6)$ % (no post-selection) to $(83 \pm 6)$ %.

In Suppl. Fig. 4d&e, we characterize the trade-off between higher charge state fidelity and readout noise: Improved charge initialization can be reached for shorter $t_{pcs}$ or higher $n_{thresh}$, but as more photons are discarded by the thresholding, the photon shot noise of the readout estimated from the total number of detected photons $n_{phots}$ as $\mathrm{SNR}_{shot}^{-1} = 1/\sqrt{n_{phots}}$ increases.

We tune the charge state by varying $n_{thresh}$ in Suppl. Fig. 4f. Here, we utilize the $\sigma_y$ component of the DEER oscillation as an alternative probe of the charge state and observe that its asymmetry reduces to $(0.12 \pm 0.12)$ when increasing to $n_{thresh} = 10$, indicating a clearly improved charge state initialization.

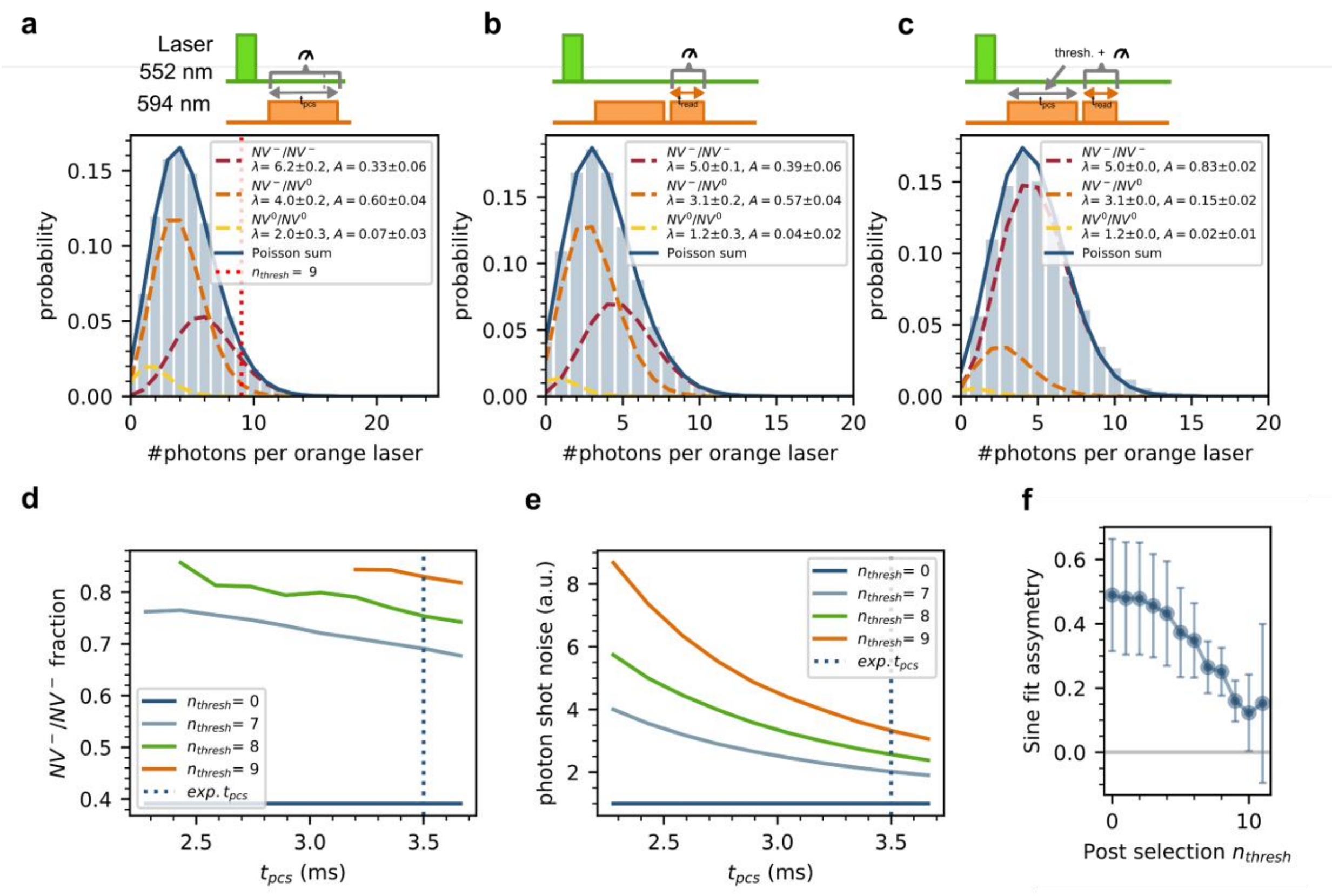

Suppl. Figure 4. Charge initialization & readout. **(a)** Histogram of photons collected during the $t_{pcs}$=3.5 ms orange charge initialization pulse. Before, the register is initialized with a 3 µs green laser pulse. Dashed lines are the single Poissonian contributions with fitted mean rates $\lambda_i$ and weight $A_i$. The chosen thresholding photon number $n_{thresh} = 9$ used for charge initialization is shown as red dotted line. **(b)** Photon histogram collected during the 2.9 ms charge readout laser pulse after the charge initialization laser. **(c)** Photon histogram with same data as (b), but after applying post-selection to analyze only charge readout data if more than $n_{thresh} = 9$ photons are detected during the charge initialization pulse. The rates $\lambda_i$ are fixed during fitting to the values extracted in (b). We conservatively estimate the error on the charge state fidelity by giving the uncertainty of the weight parameter $A_{--}$ in (b). **(d)** $NV^-/NV^-$ charge initialization fidelity measured for different length $t_{pcs}$ of the gating window which counts photons during the charge initialization laser (same data set as in (b)&(c)). A shorter $t_{pcs}$ shifts the Poissonian in the histogram to the left and thus increases charge state fidelity. Unstable fits are discarded. **(e)** Photon shot noise as calculated from the number of analyzed photons $(1/\sqrt{n_{phots}})$ and normalized to the noise without post-selection. Increasing the charge state fidelity comes at the cost of higher readout noise. **(f)** Asymmetry of the $\sigma_y$ component of the DEER oscillation presented in Fig 2e while varying the post selection threshold parameter $n_{thresh}$.

## S.VI. Spam and Entangling Gate Fidelity in a Magnetic Field

Misaligning the magnetic field against the NV axis in the diamond lattice lowers the spin initialization fidelity by green laser excitation and yields a partly mixed spin state. This effect originates from the altered effective decay rates into the new magnetic ground state levels that are superpositions of the aligned spin ground states[105]. Here, we estimate the SPAM error given in the main text caused by a reduced spin initialization.

In our (misaligned) magnetic field setting 2, a reduced readout contrast is observed for the misaligned NV 1 in Suppl. Fig. 5a. We avoid the influence of different orientations of both NVs' optical dipoles by determining the maximum readout contrast for varying orientation of the linearly polarized green (552 nm) excitation laser by a $\lambda/2$ wave plate.

In Suppl. Fig. 5b, we verify that transformation of a 7 level rate equation model[105] agrees reasonably well with the experimentally observed relative Rabi contrast $\max(contr_{NV1})$ / $\max(contr_{NV2})$ at different absolute magnetic field values. The misalignment angle $\theta \approx 74°$ of each datapoint is similar as in magnetic field setting 2, up to a precision of $\pm 2°$.

We analyze the model with two sets of rates (see Suppl. Table 2):

1. Optical rates as given in [106]. To calibrate the laser pump rate as the only free parameter, we minimize the difference to the experimental data. Comparing to our experimental data, we observe a deviation from the model.
2. Since the intersystem crossing rates are not straightforward to determine, we additionally adapt the rates from the $^3E$ level to the singlet and minimize the difference to experimental data in parallel to the laser pump rate. The resulting rates fit better to our experimental data.

The same model allows us to estimate the SPAM error introduced by a reduced spin polarization. The electron spin initialization fidelity $F_{init}$ is given by:

$$F_{init} = \frac{p_0}{p_0 + p_{-1} + p_{+1}} \tag{S41}$$

Note that the experimentally observed readout contrast at misaligned magnetic field in Suppl. Fig. 5a&b includes contributions from both the reduced spin polarization and the less efficient optical readout. To obtain the spin initialization fidelity, we determine the populations $p_{m_s}$ of the $m_s = 0, \pm 1$ levels after a 3 µs green laser and a waiting time of 1 µs.

In Suppl. Fig. 5c, we plot the simulated spin initialization fidelity from Suppl. Equation S41 and the entangling gate fidelity (predicted like in Fig. 4e) for varying magnetic field. At zero field (equivalent to no misalignment), the predicted spin initialization fidelity (defined in the spin 1 manifold, mean of both set of rates) is 77 %. This value is lower than in experimental work ($\geq$ 92 % [107],($88 \pm 4$) %[103]), indicating that the available models might underestimate the absolute degree of spin polarization.

We estimate the SPAM error due to spin mixing to 17 % by dividing the simulated value (mean of both set of rates) at our magnetic field setting 2 by the value at zero field:

$$err_{SPAM,init} = \frac{1 - F_{init}(B)}{1 - F_{init}(B = 0)} \tag{S42}$$

For small NV quantum registers, such a SPAM error by spin mixing seems acceptable, especially when a higher gate fidelity can be reached in turn. On the other hand, larger scale quantum processors will require to mitigate state mixing, as state preparation errors are only inefficiently correctable[108].

We observe a trade-off between higher gate fidelity but lower spin initialization fidelity when increasing the magnetic field. Depending on the application, diamond quantum registers with NVs of different orientations may be optimized towards higher entangling gate fidelity or reduced SPAM spin initialization error.

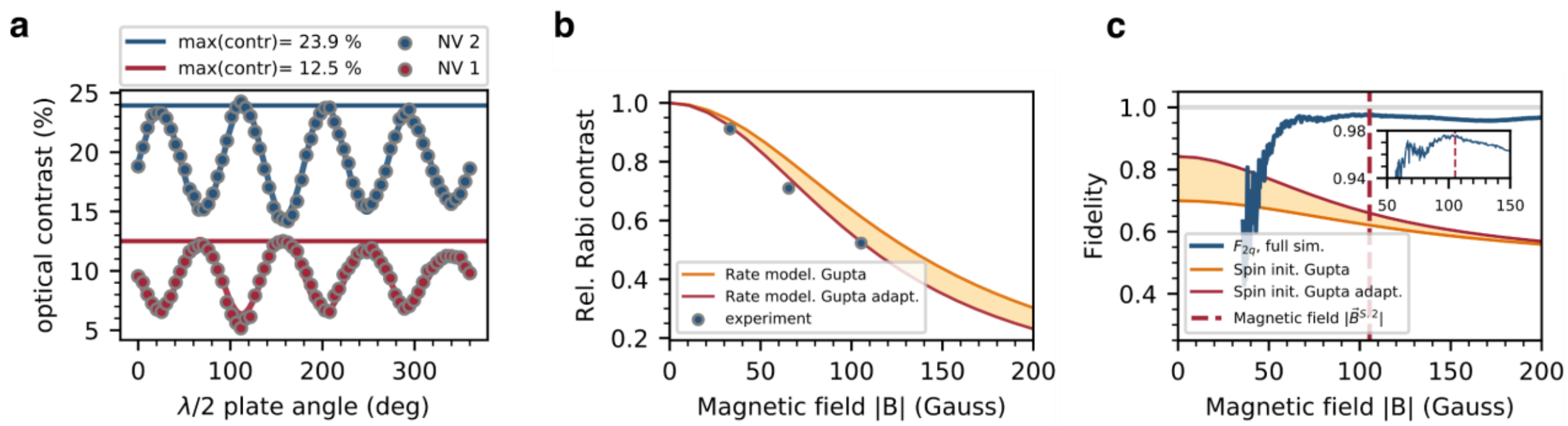

Suppl. Figure 5. SPAM error by spin mixing. (a) Optical contrast as measured from fits to Rabi oscillations on both NVs at different $\lambda/2$ wave plate angles with magnetic field setting 2. The maximum optical contrasts of both NVs is determined from the maximum value of a fitted empirical model (sum of two sines). (b) Relative Rabi contrast $\max(contr_{NV1})$ / $\max(contr_{NV2})$ at different absolute values of the magnetic field and misalignment angles $\theta_{NV1} \sim 74°$ , experimental precision $\pm 2°$. The solid line is a simulation based on the rate model and transition rates in Suppl. Table 2. Circles are measurement data taken at different magnetic fields as described in (a). (c) Simulated $\sqrt{ZZ}$ gate fidelity and spin initialization fidelity for different absolute values of the magnetic field and a misalignment angles $\theta_{NV1} = 74.08°$ , $\theta_{NV2} = 3.58°$.

Suppl. Table 2: Rates used for modelling (as in [105]) the readout contrast and spin initialization fidelity in a misaligned magnetic field.

| Rates | Model Gupta[106], NV 1 | Model Gupta, adapted ISC |
|---|---|---|
| $k_{57}$ | 92.9 MHz | **90.307 MHz** |
| $k_{67}$ | 92.9 MHz | **90.307 MHz** |
| $k_{47}$ | 11.2 MHz | **3.004 MHz** |
| $k_{71}$ | 4.9 MHz | 4.9 MHz |
| $k_{72}$ | 2.03 MHz | 2.03 MHz |
| $k_{73}$ | 2.03 MHz | 2.03 MHz |
| $k_{41}$ | 66.08 MHz | 66.08 MHz |
| $k_{52}$ | 66.08 MHz | 66.08 MHz |
| $k_{63}$ | 66.08 MHz | 66.08 MHz |
| laser pump $\beta$ | 1.215 | 1.938 |

## S.VII. Simulated Gate Fidelities

In the main text, we discuss the error sources determining the gate set fidelity. In Suppl. Fig. 6a, these errors are illustrated in the level structure of the coupled NV system. After spin initialization, both NVs are polarized into the $m_s = 0$ sublevels. Due to the unpolarized nuclear $^{14}$N spins at ambient conditions, the initialized state is a mixed state with nearly equal classical probability of the nitrogen populations. However, our quantum register is not formed from the $^{14}$N levels and initially the electron states are pure (after taking the partial trace).

After and during preparation of arbitrary register input states, multiple gate errors occur:

- As the inherent $^{14}$N spins of both NVs are unpolarized, the nuclear spin state is random at the beginning of every experimental shot. Consequently, most of the microwave pulses suffer from a detuning $\Delta = \pm A_{zz}/(2\pi) = \pm 2.16$ MHz from the correct microwave transition frequency for the electron, if the nitrogen state is $m_I = \pm 1$. This leads to imperfect driving of the three hyperfine lines, as indicated by the orange sketch of the pulse spectrum. Additionally, the unpolarized nitrogen spin causes free evolution on the non-addressed NV in our single qubit gate implementation (that we discuss in Suppl. Note 3).
- In a magnetic field setting that is aligned to the NV quantization axis, the diagonal hyperfine interaction tensor $\underline{A} = \sum_{i=x,y,z} A_{ii} S_i \otimes I_i$ in the Hamiltonian (see Methods, Suppl. Note 2) effectively only contains $S_z \otimes I_z$ terms. Here, $S_i, I_j$ denote the $S = I = 1$ spin operator of the NV and the $^{14}$N spin, respectively. In our geometry however, magnetic field misalignment is unavoidable and in the tilted basis the hyperfine tensor is no longer diagonal. Then, terms of, e.g., the form $S_z \otimes I_x$ appear and can cause entanglement with the nitrogen spin or population transfer from the qubit to the nitrogen spin. Especially during gate operations that contain dynamical decoupling pulses, this decreases the electron spin polarization (see Suppl. Note 4).
- All microwave pulses (driving $m_s = 0$,-1 on NV 2, $m_s = 0$,1 on NV 1) generate a small, unwanted microwave drive on the other NV (“crosstalk”, dashed arrows in Suppl. Fig. 6a) or transitions to states out of the qubit subspaces (“leakage”, solid arrows).

Note that all error sources may not only change populations, but also cause unwanted phase shifts.

In Fig. 4e, we predict the entangling gate fidelity of only the $\sqrt{ZZ}$ gate from our model. We use the same simulation to give gate fidelities for the single-qubit gates of the gate set in Suppl. Fig. 6b. As the single-qubit pulses in our experiment don’t employ dynamical decoupling, the fidelity improvement expected from polarized $^{14}$N nitrogen spins is more pronounced. At high enough Rabi frequencies, we expect nearly no further improvement by aligning the magnetic field. Consequently, we anticipate that polarizing the nitrogen spins and mitigating crosstalk and leakage by optimal control could enable very high-quality single-qubit control in multi-NV diamond quantum registers.

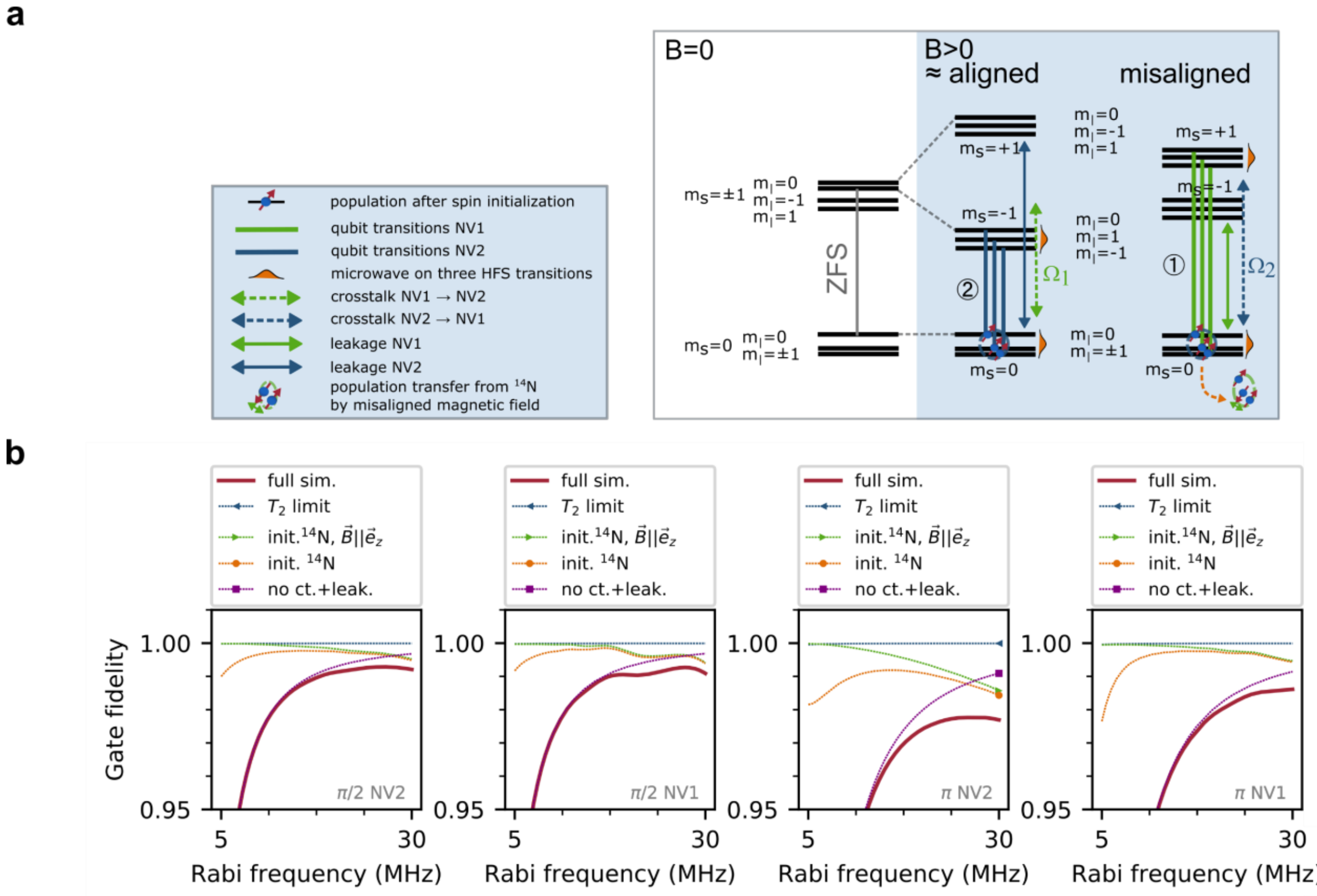

Suppl. Figure 6. Gate errors. **(a)** Level scheme of the $^3A_2$ level of two NVs including Zeeman and $^{14}$N hyperfine interaction. The error sources discussed in the main text are sketched. After spin initialization by the green laser, we obtain pure electron states indicated by blue dashed circles after tracing out the $^{14}$N nuclear spins. During a gate, interaction with the thermal nitrogen spin causes depolarization of the electron spin (green dashed ellipse). **(b)** Simulated gate fidelities for all single-qubit gates (of phase X) of the gate set for different Rabi frequencies. Decoherence is considered as $T_2$ process as described in the Methods section and much smaller for the shorter single-qubit gates.

## S.VIII. Extended Coherence Limits

Using Equation 5 with coherence times for different decoupling sequences, we estimate the coherence-limited entangling gate fidelity in realistic experimental scenarios, also cited in the Methods section F, in Suppl. Table 2.

Specifically, we note that decoherence in NV centers and other color centers in diamond can usually be modelled with an Ornstein-Uhlenbeck process[81,109,110] and one can estimate the expected memory time limit with ideal, instantaneous $\pi$ pulses for particular pulse spacing $\tau_1$ by[110]

$$\mathrm{T}_2(\tau_1) = \left(\frac{T_{2,HE}}{\tau_1}\right)^2 T_{2,HE} \tag{S43}$$

not taking $T_1$ decay into account. This results in much higher theoretical values $T_{2,\mathrm{XY8}}$ of the order of several ms for the particular $\tau_1 = 800$ ns, which is then limited by $T_1$ relaxation. The lower actual $T_{2,\mathrm{XY8}}$ in our experiment suggest that pulse errors might have an effect on the achieved decoherence times, so using robust dynamical decoupling sequences, e.g., KDD4 or the UR family[100,111] is likely to improve $T_2$. The improvement itself can be characterized in terms of the transition probability error $\epsilon$ of a single $\pi$ pulse, so the fidelity of a single XY8 sequence is[110]

$$\mathrm{F}_{XY8} \approx 1 - O(\epsilon^3) \tag{S44}$$

In comparison, the fidelity of KDD4 (KDD sequence, nested in the XY4 sequence)[111] and the UR sequences are[100,111]

$$\mathrm{F}_{KDD4} \approx 1 - O(\epsilon^5) \tag{S45}$$

$$\mathrm{F}_{URn} \approx 1 - O(\epsilon^{n/2}) \tag{S46}$$

Thus, using these advanced robust sequences could, in principle, allow to improve our coherence times towards the $2T_1$ limit, which should be of the order of milliseconds. Following the fidelity estimation in Equation 5, the decoherence limit of XY8 can be estimated to $\mathrm{F}_{2q,XY8} = 1 - pEPG_{T2} \approx 0.986$, with the decoherence time for our experiment. Improving the coherence time towards the milliseconds regime would in principle push the expected fidelity towards higher values, e.g., $\mathrm{F}_{2q,adv} \approx \exp\left(-t_{gate}/T_{2,adv}\right) \approx 0.996$ for $T_{2,adv} \approx 2$ ms with advanced dynamical decoupling sequences. We note that the assumption of simple exponential decay is conservative as the actual decay rate for dynamical decoupling with an Ornstein-Uhlenbeck noise model is typically a stretched exponential[109,110], which would result in even higher fidelities on the time scale of the gate.

Another approach for extending the coherence limit is operating at low temperatures, which can be beneficial in several aspects. First, $T_1$ relaxation due to two-phonon Raman and Orbach-type processes is negligible at temperatures of the order of 4 K, boosting $T_1$ to the order of one hour[30]. This results in several orders of improvement in the achievable coherence times e.g., $T_{2,CP,4K} \approx 1.58$ s[30]. This would in principle make the effect of decoherence negligible on the time scales of realistic NV-NV dipolar coupling, yielding $1 - \mathrm{F}_{2q,T2,4K} \approx 4 \times 10^{-6}$. Apart from negligible decoherence, operating at low temperatures in principle would allow the polarization of the nitrogen nuclear spin, reducing the need for high Rabi frequency to cover all hyperfine transitions. Operating with lower Rabi frequencies will in turn reduce the effect of crosstalk and leakage, minimizing the detrimental effect of both of these errors, as reported in Fig. 4d in the main text.

Suppl. Table 2. $T_2$ coherence times in various experimental scenarios. Assuming the coherent error sources discussed in the context of Fig. 4 are mitigated, the listed decoherence-limited entangling gate fidelities are achievable. For the extended coherence times, the same gate duration $t_{gate}$ as in the current experiment is assumed (†). The duration $t_{gate}$ of our two-qubit gate depends on the effective coupling, and considers the optimized $\tau_1 = 800$ ns.

| Scenario | Relevant coherence time | $t_{gate}$ | Decoherence-limited gate fidelity $F_{2q,T2}$ |
|---|---|---|---|
| Previous best, Room temperature, Hahn echo [24] | $\text{avg}(T_{2,\text{HE}}) = 332$ µs | 25.4 µs | 0.926 |
| Current experiment, Room temperature, Hahn echo | $\text{avg}(T_{2,\text{HE}}) = 103$ µs | 6.4 µs | 0.939 |
| Current experiment, Room temperature, XY8 | $\text{avg}(T_{2,\text{XY8}}) = 465$ µs | 6.4 µs | 0.986 |
| Advanced DD, Room temperature | $\text{avg}(T_{2,\text{aDD}}) = 2$ ms | 6.4 µs † | 0.996 |
| Carr-Purcell, Cryogenic experiment (4 K) | $\text{avg}(T_{2,\text{CP}}) = 1.58$ s [30] | 6.4 µs † | 0.999996 |

To compare our gate design with previous work, we also list the decoherence limits of a two-qubit gate under Hahn echo decoupling for both experiments in Suppl. Table 2. Due to its Hahn echo-like structure, $T_{2,HE}$ is the relevant coherence time for the gate in ref. [24]. We observe a slight improvement in the decoherence-limited gate fidelity for our sample, with an improvement factor of $(1 - F^{[24]}_{2q,T2,HE})/(1 - F_{2q,T2,HE}) = 1.21$. While the dipolar coupling is higher for our sample, the coherence time $\text{avg}(T_{2,HE})$ is shorter and the ratio $t_{gate}/\text{avg}(T_2)$ ultimately determines the fidelity limit. Our gate employs higher order decoupling, thereby extending the limiting coherence time scale to the much longer $T_{2,XY8}$. Thus our final infidelity is a factor of $(1 - F_{2q,T2,HE})/(1 - F_{2q}) = 1.53$ better than our sample's apparent $T_{2,HE}$ limit. By the increased coherence time, we access a new regime, where the error sources discussed in the main text become relevant.